\documentclass{aa}

\usepackage{epsfig,graphicx,natbib}
\usepackage{longtable}
\usepackage{amssymb}
\usepackage{amsmath}
\usepackage{mathrsfs} 
\usepackage{color}
\usepackage{subcaption}
\usepackage{booktabs}
\usepackage{xspace}
\usepackage{float}
\usepackage{longtable}
\usepackage{breqn}
\usepackage{algpseudocode}
\usepackage{algorithm}
\usepackage{hyperref}
\usepackage[raggedright]{sidecap}

\usepackage{todonotes}

\newcommand{\norm}[1]{\lVert #1 \rVert}

\newcommand{\RR}{\mathbb{R}}

\newcommand{\sgra}{Sgr~A$^*$\xspace}

\renewcommand{\textbf}[1]{#1}

\defcitealias{Issaoun2022}{Issaoun, Wielgus et al. 2022}
\defcitealias{Jorstad2022}{Jorstad, Wielgus et al. 2022}
\defcitealias{Mueller2023}{Paper I}
\defcitealias{Mus2023}{Mus (PhD Thesis)}
\defcitealias{Mus2023b}{Paper II}
\defcitealias{Zhang2008}{Zhang \& Li (2007)}
\defcitealias{Pardalos2017}{Pardalos et al. 2017}

\begin{document}

\title{Swarm intelligence for full Stokes dynamic imaging reconstruction of interferometric data}

\author{Alejandro Mus \inst{1,2,3}
\and Hendrik M\"uller \inst{4,5}\thanks{Both first authors have contributed equally to this work.}
\and Andrei Lobanov \inst{4}
}

\institute{
  Departament d’Astronomia i Astrof\'isica, Universitat de Val\`encia, C. Dr. Moliner 50, 46100 Burjassot ,Val\`encia, Spain \\
  \email{alejandro.mus@uv.es}
  \and
  Observatori Astron\`omic, Universitat de Val\`encia, Parc Cient\'ific, C. Catedr\`atico Jos\'e Beltr\'an 2, 46980 Paterna, Val\`encia, Spain
  \and
  Departament de Matem\`atiques i Inform\`atica, Universitat de les Illes Balears, E-07071 Palma de Mallorca, Spain
  \and 
   Max-Planck-Institut für Radioastronomie, Auf dem Hügel 69, D-53121 Bonn (Endenich), Germany \\ \email{hmueller@mpifr-bonn.mpg.de}
   \and Jansky Fellow of National Radio Astronomy Observatory, 1011 Lopezville Rd, Socorro, NM 87801, USA
}

\date {Received  / Accepted}

\authorrunning{Mus+M\"uller}
\titlerunning{Swarm Intelligence for reconstructing interferometric data}

\abstract
{In very long baseline interferometry (VLBI) the combination of \textbf{multiple antennas permits the synthesis of a virtual telescope with a larger diameter and consequently higher resolution than the individual antennae}. Yet, due to the sparse nature of the array, \textbf{recovering} an image from the observed data is a challenging ill-posed inverse problem.}%
{The VLBI community is interested in not only recovering an image in total intensity from interferometric data, but also to obtain results in the polarimetric and the temporal domain. Only a few algorithms are able to work \textbf{in all these domains simultaneously}. In particular, the algorithms based on optimization that \textbf{consider} various penalty terms specific to static total intensity imaging, time-variability and polarimetry are restricted to \textbf{grids} the domain of the objective function. In this work we present a novel algorithm, multiobjective particle swarm optimization (MO-PSO), that is able to recover the optimal weights without any space-gridding, and to obtain the marginal contribution of each the playing terms.}
{To this end, we utilize multiobjective optimization together with particle swarm metaheuristics. We let the swarm of weights to converge together to the best position.}
{We \textbf{evaluate} our algorithm with synthetic data sets that are representative for the main science targets and instrumental configuration of the Event Horizon Telescope Collaboration (EHTC) and its planned successors. We successfully recover the polarimetric, static and time-dynamic signature of the ground truth movie, even with relative sparsity, and a set of realistic data corruptions.}
{We have built a novel, fast, hyperparameter space gridding-free algorithm that successfully recovers static and dynamic polarimetric reconstructions. Compared to Regularized Maximum Likelihood (RML) methods, it avoids the need for parameter surveys, and it is not limited to the number of pixels such as recently proposed multiobjective imaging algorithms. Hence, this technique is a novel, useful alternative tool to characterize full stokes time-(in)dependent signatures in a VLBI data set robustly with a minimal set of user-based choices.}

\maketitle

\section{Introduction}

Very Long Baseline Interferometry (VLBI) \textbf{utilizes} an array of multiple antennas to achieve unparalleled astronomical resolutions. Within this array, each antenna concurrently observes a celestial source, following the \textbf{procedure} outlined in the Zernike-van Cittert theorem~\citep{Thompson2017}. The correlated signals from antenna pairs, in accordance with this theorem, approximate the Fourier transform of the true brightness distribution of the sky. The baseline distance between antennas on the celestial plane determines the specific Fourier frequency.

In an ideal scenario, a fully sampled Fourier domain ("$u,v$ plane") permits the retrieval of the original image via a simple inverse Fourier transform. Practical limitations like a limited number of antennas, restricted observing time, noise, and instrumental effects lead to sparse Fourier coefficient measurements. This subset, denoted as "$u,v$-coverage," represents observed spatial Fourier frequencies. \textbf{The above corruptions render the image reconstruction problem ill-posed.}

Global VLBI, especially at millimeter wavelengths, introduces additional hurdles like a smaller number of antennas, imprecise visibility calibration, and lower signal-to-noise ratios compared to longer wavelength arrays. Consequently, the reconstruction problem lacks strong constraints, displaying non-convex and possibly multimodal traits due to missing data.
Successful imaging hinges on incorporating robust prior information.

Traditionally, CLEAN and its variations \citep{Hogbom1974, Clark1980, Bhatnagar2004, Cornwell2008, Rau2011, Offringa2014} were the go-to methods for imaging. However, recent years have witnessed the emergence of new imaging algorithms for global VLBI data tailored to the Event Horizon Telescope's (EHT) needs. \textbf{These additional methods fall mainly into} four categories: super-resolving CLEAN variants \citep{Mueller2022b}, Bayesian methods \citep{Arras2022, Broderick2020, Tiede2022}, regularized maximum likelihood (RML) methods \citep{Chael2016, Chael2018, Akiyama2017, Akiyama2017b, Mueller2022}, and multiobjective optimization techniques such as MOEA/D \citep[][Paper I and Paper II hereafter]{Mueller2023,Mus2023b}.

Any of these algorithmic approaches has multiple benefits and disadvantages. In this manuscript we design a novel imaging algorithm, multiobjective particle swarm optimization (MO-PSO), that combines the speed, and accuracy of RML techniques with the high degree of automatization (i.e. a small number of hyperparameters) and global exploration of the objective landscape inherent to MOEA/D \citepalias{Zhang2008, Mueller2023, Mus2023b}. To this end we combine a particle swarm optimization algorithm in a multiobjective setting with gradient-based convex optimizations in a forward modeling (RML) framework. 

Compared to RML algorithms, MO-PSO models the sample of non-dominated, locally optimal solutions identified by Pareto optimality\textbf{,} akin to parameter surveys common for RML methods. It selects the objectively best image in an unsupervised, automatized way. MO-PSO determines optimal weight combinations without a grid search, by solving a convex problem (thus unique solution) aligning its complexity with standard RML methods, and orders of magnitudes faster than parameter survey strategies. As a byproduct, MO-PSO allows the characterization of images based on each regularizer's value.

Compared to multiobjective optimization algorithms, i.e. MOEA/D, MO-PSO solves multiple numerical limitations: 1) a discretization of the weighting space is needed for MOEA/D, 2) to improve the numerical genetic operations performance, a \textit{ad-hoc} rescaling has to be set and 3) its computational complexity increases with denser meshes and the addition of new functionals (\textbf{due to the exploration of a higher dimensional space}) in MOEA/D. These limitations are mainly caused by the evolutionary optimization strategy inherent to MOEA/D which leads to sub-optimal numerical scaling. By replacing the genetic minimization with fast inner RML imaging iterations, MO-PSO overcomes these limitations.

Moreover, MO-PSO combines these advantages without casting the algorithmic framework into a less flexible setting. Thus, it is rather straightforward to extend the algorithm to full polarimetry, dynamic reconstructions and finally the reconstruction of polarimetric movies following the strategy described in \citepalias{Mus2023b}. The capability to recover polarimetric movies in a self-consistent forward modeling framework is a unique capability shared by only few algorithms \citep{Mueller2022c, Mus2023b}. In this manuscript we discuss the utilization of MO-PSO for polarimetric movie reconstructions as well, and demonstrate its performance in synthetic data. This work offers a swift, low-complexity, self-calibration-independent, and unsupervised alternative, working in full Stokes static and time dynamic reconstructions within a unified framework, and combining and advancing several of the advantages and capabilities of recently proposed imaging algorithms.

We outline the basic concept of the algorithm in detail, but restrict the demonstration of the performance of MO-PSO to a variety of exemplary, outstanding, studies \textbf{as} the reconstruction of real 2017 EHT data and \textbf{the reconstruction of polarimetric and dynamic synthetic EHT and EHT+ngEHT data.} More examples can be found in the git specified in Sec.~\ref{sec:softwareAvailability}.

\section{VLBI measurements and polarimetry}\label{sec:vlbibases}

In this section we present a general overview of VLBI full Stokes measurements. The section presents standard radio interferometry data products and is rewritten from Sec. 2 in ~\citepalias{Mus2023b}. 

The correlated signal of an antenna pair, denoted as the visibility $\mathcal{V}(u,v)$, is described by the van Cittert-Zernike theorem~\citep{Thompson2017}:
\begin{align}
    \mathcal{V} (u, v) = \int \int I(l, m) e^{-2 \pi i (l u + m v)} dl dm \,,  \label{eq: vis}
\end{align}
This equation \textbf{states} that the true sky brightness distribution, represented by $I(l, m)$, and the observed visibilities form a Fourier pair. The coordinates $(u, v)$ are defined by the baselines that separate each antenna pair relative to the sky plane. However, it is important to note that, \textbf{in practice,} not all Fourier frequencies are measured, resulting in a sparse coverage in the $(u, v)$-plane, \textbf{and particularly sparse in VLIB}. This sparsity makes the task of imaging, which involves recovering the sky brightness distribution from the observed visibilities, an ill-posed inverse problem since the codomain is sparsely sampled. Furthermore, the visibilities are subject to calibration issues and thermal noise. To address these challenges, we must correct for calibration effects. Direction-independent calibration effects are incorporated into station-based gain factors, denoted as $g_i$. The measured visibilities for a given time $t$ observed along one antenna pair, $i,j$, are given by:
\begin{align}
    V(i,j,t) = g_{i,t} g_{j,t}^{*} \mathcal{V}(i,j,t) + N\left(i,j,t\right)\,,
\end{align}
with station-based gains $g_i, g_j$ and a baseline-dependent random noise term $N\left(i,j,t\right)$.

The light's polarized fraction recorded at each antenna is split into various polarizations by polarization filters (orthogonal linear or right/left-handed circular), contributing to four Stokes parameters: $I$, $Q$, $U$, and $V$. These parameters satisfy the inequality:
\begin{align}
 I^2 \geq Q^2+U^2+V^2, \label{eq: stokes_inequality}
\end{align}
signifying that the total intensity of polarized and unpolarized emission always surpasses the polarized intensity. Similar to Stokes $I$ imaging, Stokes visibilities for different parameters are identified as~\citep{Thompson2017}:
\begin{align}
    &\mathcal{V}_I =  \mathcal{F}I, \\
    &\mathcal{V}_Q =  \mathcal{F}Q, \\
    &\mathcal{V}_U =  \mathcal{F}U, \\
    &\mathcal{V}_V =  \mathcal{F}V,
\end{align}
being $\mathcal{F}$ the Fourier transform functional.
The Stokes visibilities are considered the \textbf{observables} for the remainder of this manuscript. \textbf{Polarimetry also suffers from station-based gains and thermal noise across all bands, alongside dealing with leakages and feed rotations. Particularly, the leakage between the perpendicular polarization filters introduces cross-terms \citep{Thompson2017}.}

\section{Multiobjective optimization}

Multiobjective optimization is a branch of mathematics that deals with the solution of optimizing a set of (\textbf{potentially} contradicting) objective functions at the same time~\citep[for instance, see][]{Pardalos2017}.
Given a vector functional, $F:=\left(f_1,\ldots,f_n\right)$, the multiobjective optimization problem (MOP) is 

\begin{problem}[MOP formulation]
  \begin{equation*}
  \label{prob:mop_ours}
  \tag{$\text{MOP}$}
    \begin{aligned}
      & \underset{x\in D}{\text{min}}
      & & F:=\left(f_1\left(x\right),,\ldots,f_n\left(x\right)\right),\\
      & \text{subject to}
      & & \mathbf{x} \in \mathbf{F} \subseteq \mathbb{R}^p.\\
    \end{aligned}
  \end{equation*}
\end{problem}

We define optimality in such a multiobjective setting by Pareto optimality: A solution is called Pareto optimal if the further minimization along one objective functional $f_i$ automatically has to worsen the scoring in another one \citep{Pardalos2017}. The set of all Pareto optimal solutions is called the Pareto front.

Several strategies \textbf{exist} for solving this kind of problem. In particular,~\citepalias{Mueller2023, Mus2023b} presented a technique using the Multiobjective Evolutionary Algorithm by Decomposition (MOEA/D)~\citep{Zhang2008} for obtaining static, and dynamic Stokes I and polarimetric reconstructions of interferometric data.

In this work, we scalarize the vector function $F$ in the form of the scalarized MOP formulation:

\begin{problem}[Scalarized MOP formulation]
  \begin{equation*}
  \label{prob:mop_moead}
  \tag{$\text{MOP-Scalarized}$}
    \begin{aligned}
      & \underset{x\in D}{\text{min}}
      & & \displaystyle{\sum_{i=i}^{n}}\ \lambda_i f_i\left(x\right),\\
      & \text{subject to}
      & & \mathbf{x} \in \mathbf{F} \subseteq \mathbb{R}^p.\\
    \end{aligned}
  \end{equation*}
\end{problem}

Let us illustrate the conceptual idea behind this scalarization. If the various objectives are contradicting to each other, the various objectives promote different parts of the solution space. In this way, we can survey local minima by different configurations of the $\lambda_i$ vectors (i.e. by different scalarizations). The sample of solutions to each of these scalarized problems approximates the Pareto front as long as the (normalized) weighting arrays $\lambda_i$ span the complete range of possible configurations.

A particular solution vector of the MOP is the \textit{ideal vector}, formed by 
\begin{equation}
    z_i = \inf f_i\left(x\right),\ \text{for } i = 1,\ldots, n.
\end{equation}

The ideal determines a lower bound of the Pareto front. The distance of one Pareto optimal solution to the ideal defines a metric that can be used as a selection criterion. It has been demonstrated in ~\citepalias{Mueller2023,Mus2023b} that the Pareto optimal solution that is closest to the ideal point represents the objectively best reconstruction within the Pareto front for VLBI (static, dynamic, and polarimetric) imaging problems.

\section{Particle Swarm Optimization}

Particle Swarm Optimization (PSO) \textbf{is a population-based stochastic optimization algorithm used to find the optimal solution to a problem by iteratively improving a swarm of candidate solutions (particles) according to a specified fitness function.}
The fundamental basis of the socio-cognitive learning process within PSO lies in a particle's individual experience and its observation of the most successful particle.

A basic PSO algorithm~\citep[see][for an extensive explanation of this algorithm and its variants]{Du2016} could be described as follows. In an optimization scenario with $n$ variables, a swarm of $N_p$ particles is established, each assigned a random $n$-dimensional position representing a potential solution. Every particle possesses its trajectory - the position ($w_i$) and velocity ($v_i$) - moving in the search space via sequential updates.

The particles of the swarm adapt their trajectories by referencing the best positions they or others have previously visited. All particles possess fitness values evaluated by an optimization-focused fitness function. The particles navigate the solution space by following the currently optimal particles. The algorithm initiates a set of particles with random positions and proceeds to explore optima through successive generations.

In each iteration, particles are updated by considering two significant values: their individual best, indicated as $w^{*}_{i}$ for $i = 1,...,N_p$, representing their personal best solution achieved thus far, and the global best, denoted as $w^{g}$, signifying the best value obtained by any particle within the population. 

At iteration $t + 1$, the swarm undergoes an update by

\begin{align}
    & v_i (t + 1) = k v_i (t) + c_1r_1 \left[w_{i}^{*} (t) - w_i (t)\right] + c_2r_2 \left[w^{g}(t) - w_i (t)\right].\label{eq:velocity_particle}\\
    & w_i (t + 1) = w_i (t) + v_i (t + 1),\  i=1,\ldots, N_p.\label{eq:position_particle}
\end{align}
PSO finds a global solution by adapting the velocity vector of each particle according to its personal best (cognition aspect, $c_1$) and the global best (social aspect, $c_2$) positions of particles in the entire swarm at each iteration. An inertia $k$ can be imposed to accelerate the convergence, and $r_1, r_2$ are two random numbers drawn from $\mathcal{N}\left(0,1\right)$.

\section{Modelization of the Problem}
We search for the Pareto optimal solution that is closest to the ideal point following the strategy presented in \citepalias{Mueller2023, Mus2023b}. Hence, we minimize the functional:
\begin{equation}
    J(x):=\lVert \left(z_1-f_1\left(x^{\textbf{p}}\right),\ldots,z_n-f_n\left(x^{\textbf{p}}\right)\right)\rVert^2,
\end{equation}
where $z = (z_1, z_2, ...)$ is the ideal point, and $x^{\textbf{p}}$ is the solution to the scalarized problem \eqref{prob:mop_moead} with respect to the weight vector $\lambda = w^{\textbf{p}}$. We aim to optimize for the best weight vectors $w^{\textbf{p}}$ \textbf{by PSO (where every particle is a vector of weights $w^{\textbf{p}}$}, the best image $x^{\textbf{p}}$ follows as a byproduct \textbf{which is evaluated using the limited memory BFGS algorithm \citep{BFGS}}.

The method is described in Algorithm~\ref{alg:shapso}. First, we initialize the PSO algorithm, i.e. the initial positions and the velocities of the particles. We can pre-compute ideal points $z_i$ associated to the objectives.

Second, we initialize a swarm of $m$ particles, $\mathbf{c^{m}_{0}}$. Each particle is a vector in $\RR^n$ and will be a combination of $\lambda_i$, i.e., the swarm is an array of weights that will be moving towards the optimal weight combination. \textbf{We iteratively solve} Prob.~\eqref{prob:mop_ours} defining the goodness of the particle via the fitness function $J$. In this way, we make the swarm move (or the weighting vectors) towards the objective closest to the ideal vector. The swarm is moved by updating the velocity and position as described by Eq.~\eqref{eq:velocity_particle} and Eq.~\eqref{eq:position_particle}. The movement of a single particle will be accepted once the fitness function $J$ improves. Note that for this evaluation we have to find the optimal solution to problem \eqref{prob:mop_moead} which is a second, internal optimization problem.

Finally, once the swarm has converged, the swarm of particles $w^p$ and their corresponding evaluations $J^p$ constitutes a sample of local minima based on the geometry of the Pareto front. To find the global minimum, we select the particle with the smallest distance to the ideal point. We evaluate Prob. \eqref{prob:mop_moead} with the globally best weight combination $w^*$. The result is the primary product of the MO-PSO algorithm: A good approximation to the on-sky intensity distribution.

The external minimization is done by PSO, the internal minimization \textbf{(which is equivalent to the classical RML imaging problem)} by BFGS optimization. In this sense MO-PSO combines multiobjective, evolutionary algorithms with RML imaging.

\begin{algorithm}
\caption{Multiobjective PSO}\label{alg:shapso}
    \begin{algorithmic}
        \Require The set of objective-players of the MOP to be solved $\left(f_1,\ldots,f_n\right)$
        \Require Bounds on RML weights: $b_{lo}, b_{up}$\\
        
        \State Create a swarm of particles $\bar{w}^{p}$, each one being a vector of $\RR^n$, with positions drawn from a uniform distribution $s \sim U(b_{lo}, U_{up})$ and velocities following a Normal distribution $v \sim\mathcal{N}\left(0,1\right)$\footnote{The initial distribution can change, if the user desires.}
        
        \State Find ideal point: $\bar{F}_i \gets \min_x f_i(x)$ for $i=1,2, ...,n$\\
        
        \While{termination criterion is not met}
            \For{each particle, $w^p$,  of the swarm}
                \State Update particle velocities $v^p$ following Eq.~\eqref{eq:velocity_particle}
                \State Update positions $\bar{w}^p$ via Eq.~\eqref{eq:position_particle}
                \State $s^{p}\gets \mathrm{argmin}_x\left\lbrace{\sum_i \bar{w}^{p}_{i}f_i\left(x\right)}\right\rbrace$
                \State $F^p \gets (f_1(s^p), f_2(s^p), ..., f_n(s^p))$
                \State $\bar{J}^p \gets \norm{\bar{F}-F^p}^2$
                \If{$\bar{J}^p < J^p$}
                    \State $J^p \gets \bar{J}^p$
                    \State $w^p \gets \bar{w}^p$
                \EndIf
            \EndFor
        \EndWhile\\
        
        \State Select the $w^p$ with lower value of $J^p$, denoted by $w^*$
        \State Solve $\text{PSOimage}\gets \arg\min_x\left\lbrace{\sum_i w^* f_i\left(x\right)}\right\rbrace$
    \end{algorithmic}
\end{algorithm}

\section{Functionals} \label{sec: functionals}

In this section, we present the different functionals used, in increasing (problem) complexity. \textbf{We first start by solving the} total intensity problem. \textbf{We then model the polarimetry}, followed by dynamic imaging and \textbf{conclude} by the polarimetric dynamic problem.

\subsection{Total intensity}

Reconstructing static total intensity images is a well-studied problem. In the context of interferometry using forward modeling, there are two main families~\citep[see][for a more extended review]{MusThesis}: 1) Backward modeling~\cite{Hogbom1974,Cornwell2008,Mueller2022b} and forward modeling~\citep[][for instance]{Honma2014, Chael2016, Akiyama2017, Akiyama2017b, Chael2018, Arras2019, eht2019d, Broderick2020, Arras2021, Mueller2022, Tiede2022}. Forward modeling algorithms group several families. For this work, we focus on RML and multiobjective formulations~\citepalias[see][]{Mueller2023, Mus2023b}. In both cases, the methods try to minimize a (set of) objective(s) including data fidelity terms that measure the fidelity of the guess solution to the observed data \textbf{and regularizers which impose conditions on the pixel brightness distribution of the image}.

\subsubsection{Data fidelity terms}
Given the number of observed visibilities $N_{vis}$, visibilities of \textbf{the} recovered solution $\mathscr{V}_i$ and error $\Sigma_i$, the fit to the visibilities is given by:
\begin{align}
    S_{\text{vis}} (\mathscr{V}, V) = \frac{1}{N_{\text{vis}}} \sum_{i=1}^{N_{\text{vis}}} \frac{|\mathscr{V}_i - V_i|^2}{\Sigma_i^2}\,,
\end{align}
and to the amplitudes by
\begin{align}
    S_{\text{amp}} (\mathscr{V}, V) = \frac{1}{N_{\text{vis}}} \sum_{i=1}^{N_{\text{vis}}} \frac{(|\mathscr{V}_i| - |V_i|)^2}{\Sigma_i^2}\,.
\end{align}
More complex relations, such as closures~\citep[terms independent of antenna corruptions][]{Thompson2017} can by modeled using
\begin{align}
    S_{\text{cph}} (\mathscr{V}, V) = \frac{1}{N_{\text{cph}}} \sum_{i=1}^{N_{\text{cph}}} \frac{|\Psi_i(\mathscr{V}) - \Psi_i(V)|^2}{\Sigma_{\text{cph},i}^2}\,, \label{eq: cph}
\end{align}
\begin{align}
    S_{\text{cla}} (\mathscr{V}, V) = \frac{1}{N_{\text{cla}}} \sum_{i=1}^{N_{\text{cla}}} \frac{|\ln A_i(\mathscr{V}) - \ln A_i(V)|^2}{\Sigma_{\text{cla},i}^2}\,, \label{eq: lca}
\end{align}
to fit phases ($\Psi$) and amplitudes ($A$) respectively.

\subsubsection{Regularizers}

Regularization terms measure the feasibility of the solution to fit the data with a model that is as simple as possible. Usual choices include are:
\begin{align}
    &\text{Total flux: } R_{\text{flux}}(I, f) = \norm{\int \int I(l,m)dldm-f}\,, \\
    &l_1 \text{-norm: } R_{l1}(I) = \norm{I}_{l^1}\,, \\
    &l_2 \text{-norm: } R_{l2}(I) = \norm{I}_{l^2}\,, \\
    &\text{Total variance: } R_{tv}(I) = \int \norm{\nabla I}\ dl\ dm \,, \\
    &\text{TV squared: } R_{tsv}(I) = \sqrt{ \int \norm{\nabla I}^2\ dl\ dm }\,, \\
    &\text{Entropy: } R_{\text{entr}}\left(I\right) = 
        \displaystyle{\int \int} I \ln\left(\dfrac{I}{M}\right) dl dm\,,
\end{align}

where $M$ denotes the brightness distribution of a model image \textbf{and} $f$ is the total flux~\citep{Akiyama2017,Chael2018,Mueller2023}.\\

In the standard formulation of these terms, they are added with corresponding weighting parameters $\alpha,\beta,...,\kappa \in \mathbb{R}$ to create the full objective functional $\mathscr{F}$:
\begin{multline}
    \mathscr{F} = \alpha S_{\text{vis}} + \beta S_{\text{amp}} + \gamma S_{\text{clp}} + \delta S_{\text{cla}} \\
    + \epsilon R_{\text{flux}} + \zeta R_{l1} + \eta R_{l2} + \theta R_{tv} + \iota R_{tsv} + \kappa R_{\text{entr}}. \label{eq: rml}
\end{multline}

In this work, we developed a novel approach, able to find the marginal contribution of each regularizer without performing any survey. To this end, we reconstruct an image in total intensity considering the seven objective functional:

\begin{align}
    & f_1:= \zeta R_{l1},\\
    & f_2:=\theta R_{tv},\\
    & f_3:=\tau R_{tsv},\\
    & f_4:=\eta R_{l2},\\
    & f_5:=\epsilon R_{flux},\\
    & f_6:=\kappa R_{entr},\\
    & f_7:=\alpha S_{vis}+\beta S_{amp}+\gamma S_{clp}+\delta S_{cla}.\\
\end{align}

For the remainder of this manuscript\textbf{,} we will restrict ourselves to $\alpha = \beta = 0$, i.e. we study calibration independent imaging from the closure products only. In this way, we make the Stokes I imaging independent of the self-calibration procedure, at the cost of an increased data sparsity, i.e. the number of statistically independent closure quantities is smaller than the number of statistically independent visibilities. However, it has been demonstrated in recent works that in the setting of the VLBI, closure quantities are constraining enough to retrieve the image structure in the presence of strong prior information \citep{Chael2018, Mueller2022, Arras2022, Thyagarajan2023}. Particularly, in \citepalias{Mueller2023}, we have shown that the degeneracies inherent to closure-only imaging are effectively \textbf{casted} into the Pareto front. This indicates that multiobjective evolution provides a reasonable framework to deal with closure-only imaging as proceeded for the remainder of this manuscript.

\subsection{Polarimetry}\label{subsec:Polarimetry}

\subsubsection{Data term}

In the context of polarimetry, we incorporate a $\chi^2$-fit approach to evaluate the linear polarized visibilities, denoted as $\mathcal{V}_Q$ and $\mathcal{V}_U$. This incorporation is made as an additional component in the form of a data term denoted by $S_{pvis}$:
\begin{equation}
    S_{pvis} (\mathscr{V}^P, V^P) := \dfrac{1}{2N_{\mathrm{vis}}}\displaystyle{\sum_{i=1}^{N_{\mathrm{vis}}}}\dfrac{\lvert \mathscr{V}^P_{i} - V^P_{i}\rvert}{\Sigma^{2}_{P,i}}.
\end{equation}
\textbf{Here $\mathscr{V}^P = \mathscr{V}_Q+i\mathscr{V}_U$ denotes the visibility of the complex linear polarization vector $P=Q+iU$, and $V^P$ and $\Sigma_P$ are the corresponding model visibilities and errors.}

\subsubsection{Regularizers}

We have used the same regularizer functionals appearing in~\cite{Mus2023b} and defined in \citet{Chael2016} and that have proven to work as shown in~\citep[e.g.][]{eht2021a}. In particular, we make use of the functionals known as $R_{ms}$, $R_{hw}$, and $R_{ptv}$, which we will describe in more detail in the subsequent sections.

First, we use the conventional KL polarimetric entropy~\citep{Ponsonby1973,Narayan1986,Holdaway1988,Chael2016} of an image $I$ with respect to a model $M_i$:
\begin{align}
    R_{hw} := \displaystyle{\sum^{N^2}_{i}} I_i \left( \ln(\dfrac{I_i}{M_i e}) + \dfrac{1+m_i}{2} \ln\left(\dfrac{1+m_i}{2}\right) \right.\\
    \left. + \dfrac{1-m_i}{2} \ln\left(\dfrac{1-m_i}{2}\right)\right). \label{eq:shw}
\end{align} 
Here $m_i$ denotes the fraction of linear polarization in pixel $i$. $R_{hw}$ favors images of $N$ pixels with fractional polarization, $\Vec{m}$ less than one.

The second regularizer $R_{ms}$ is:
\begin{equation}\label{eq:msimple}
    R_{ms} = \displaystyle{\sum^{N}_{i}}\lvert I_i\rvert \ln \lvert m_i\rvert
\end{equation}
Finally, we utilize the polarimetric counterpart of the total variation regularizer \citep{Chael2016}:
\begin{equation}\label{eq:tv}
    R_{ptv} = \sum_{i}\sum_{j}\sqrt{\vert| P_{i+1, j} - P_{i,j} \rvert^2 + \left| P_{i, j+1} - P_{i,j} \right|^2},
\end{equation}
where $P$ is the complex realization of the linear polarized emission, i.e. $P = Q+iU$.

In conclusion, the resulting multiobjective problem consists of the single functionals: 
\begin{align}
    & f_1:=\beta R_{ms}, \label{eq: pol1}\\
    & f_2:=\gamma R_{hw},\label{eq: pol2}\\
    & f_3:=\delta R_{ptv},\label{eq: pol3}\\
    & f_4:=\alpha S_{pvis}. \label{eq: pol4}
\end{align}

\subsection{Dynamics}\label{subsec:Dynamics}

Dynamics are modeled using the ngMEM functional~\citep{Mus2023b, Mus2023}.
Here we present a brief formulation\textbf{,} and we refer to the bibliography for further details.
We consider a time partition of the observation times into $r$ keyframes. The visibilities of one frame are:
\begin{equation}\label{eq:keyframes}
\mathcal{V}^l = \left\{\mathcal{V} ~\mathrm{for}~~ t \in [t_l - \Delta t_l/2, t_l + \Delta t_l/2] \right\}, ~~ l \in [1,r],
\end{equation}
where $t^l$ is the observing time, and $\Delta t_l$ is a frame-dependent scalar that determines its duration. For each $l$, the associated data set $\mathcal{V}^l$ produces an ``image keyframe''. The model will have a total of $r \times N^2$ parameters (i.e., $r$ images of $N^2$ pixels each).

\subsubsection{Data term}

Now we naturally extend the data terms and penalty terms to a time-series, e.g. by:
\begin{align}
S_{vis}(\mathcal{V}, p) = \frac{1}{r} \sum_{l=1}^r S_{vis}(\mathcal{V}^l, p^l),
\end{align}
where $p$ is a time-series of images (i.e. a movie). We proceed analogously for all the other data terms and regularization terms.

\subsubsection{Regularizer}

In~\citetalias{Mus2023b}, we modeled the dynamical imaging problem by including the $f_{\mathrm{ngMEM}}$ functional. This regularizer, introduced in~\citep{MusThesis, Mus2023} acts on a sequence of images $p \in \RR^{r \times N^2}_{+}$: 
\begin{equation} \nonumber
\mu_{\mathrm{ngMEM}}R_{\mathrm{ngMEM}} = \sum_{n, j \neq k}{T_n^{jk}},
\end{equation}
where  $j,k \in \{1, 2, ..., r\}$ denote the frame in the time-series, $\mu_{\mathrm{ngMEM}}$ is a hyper-parameter and
\begin{equation}
\label{eq:image_memory}
    T_n^{jk} := 
    e^{-\frac{|t^j - t^k|^2}{2\sigma^2}}
    \left(|p_n^j - p_n^k |+C\right)\log{\left( |p_n^j - p_n^k| + C\right)}.
\end{equation}
Here, $t^j$ denotes the times of the associated keyframe $j$ and $\sigma$ and $\mu_{\mathrm{ngMEM}}$ are two a priori hyper-parameters. While $\sigma$ has a minor effect, $\mu_{\mathrm{ngMEM}}$ is the weight \textbf{considered} as a particle of our swarm. We refer to~\citep{Mus2023b, Mus2023} for further details.

We can solve this problem of dynamic imaging (i.e., find the set of image keyframes that optimally fit the data) by using the following formalism. 
\begin{align}
    & f_1:=\zeta R_{l1}, \label{eq: dyn1}\\
    & f_2:=\theta R_{tv},\label{eq: dyn2}\\
    & f_3:=\tau R_{tsv},\label{eq: dyn3}\\
    & f_4:=\eta R_{l2},\label{eq: dyn4}\\
    & f_5:=\epsilon R_{flux},\label{eq: dyn5}\\
    & f_{6}:=\kappa R_{\mathrm{entr}}, \label{eq: dyn6}\\ 
    & f_7:=\mu R_{\mathrm{ngMEM}}, \label{eq: dyn7} \\
    & f_8:=\alpha S_{vis}+\beta S_{amp}+\gamma S_{clp}+\delta S_{cla}.\label{eq: dyn8}
\end{align}

\subsection{Dynamic polarimetry}

To solve the dynamic polarimetric problem, \textbf{one can adopt} two strategies.

\textbf{The first strategy} can be to emulate the \textbf{procedure} presented in~\citetalias{Mus2023b} for MOEA/D and \citet{Mueller2022c} for DoG-HiT, intended to avoid high-dimensional problem\textbf{s} of the Pareto front. This strategy consists in splitting into several steps. First, we solve for the dynamics in total intensity as described \textbf{in Sec.~\ref{subsec:Dynamics}}. We self-calibrate the data set and do the leakage calibration. Second we calculate a static polarimetric image as described \textbf{in Sec.~\ref{subsec:Polarimetry}}. Third, we cut the observation into keyframes and solve the polarimetric imaging at every keyframe independently using Algorithm~\ref{alg:shapso}, initializing the population with the final population of the static polarization step and assuming the Stokes I image from the time-dynamic exploration.

\textbf{The second strategy} could \textbf{be} add all polarimetric, time-dynamic and static data terms together to the same multi-objective problem. While the problem formulation would be straightforward, and the numerical complexity acceptable, this strategy does not represent the strategy that will be applied in practice since a significant part of the forward is still missing: We do not include the leakage terms in the data terms. One way to circumvent this difficulty would be to use the \textbf{properties} of closure traces \citep{Broderick2020b} directly for leakage-independent imaging. However, this approach comes with its own challenges, and is deferred to a consecutive work. For the remainder of this manuscript, we emulate the strategy presented in \citetalias{Mus2023b}.

\section{Comparing MO-PSO reconstruction philosophy with MOEA/D and single-objective philosophies}

In this section, we compare and remark \textbf{on} the differences between the well-known single-objective forward method, as the one presented in \texttt{ehtim}~\citep{Chael2016,Chael2018} or \texttt{smili}~\citep{Akiyama2017,Akiyama2017b}, the multi-objective imaging reconstruction~\citepalias{Mueller2023, Mus2023b} and the \textbf{method} presented here.

In general, the novel approach of MO-PSO can be summarized in \textit{seek the best hyper-parameter combination, and then check if its associated image is good.} In other words, the main effort of this algorithm \textbf{focuses} on solving the convex weight problem, and the reconstruction is a byproduct.

\subsection{MOEA/D} \label{ssec: comp_to_moead}
MOEA/D and MO-PSO share \textbf{many} similarities. They are both multiobjective evolutionary algorithms that share the same set of single-objectives. They both utilize the concept of Pareto optimality to identify the best reconstructions, and are based on the same concept of finding the Pareto-optimal solution that is closest to the ideal. There are however significant differences.

The MOEA/D formulation seeks the images which are the best compromises between the different regularizers and are closer to the model. It identifies these solutions by approximating the full Pareto front. MO-PSO also explores the Pareto front globally, but converges to the image with best possible weight combination without storing or representing the full front. We can \textbf{nevertheless} identify the robustness of solutions by MO-PSO due to its randomized evolution, see Sec. \ref{sec: reg_characterization} for more details. While MOEA/D compares each regularizer with respect to the data terms, MO-PSO \textbf{frees} each regularizer free and is able to compute a marginal distribution of each regularizer in this way. MOEA/D looses the marginal contribution of each regularizer, and \textbf{returns the} relative importance of the hyper-parameters instead.

MOEA/D explores all weight combinations. The number of combinations is given by the size of the grid (and, therefore, the number of sub-problems in which it is decomposed). In contrast, MO-PSO let the weights free and explores a ``continuous'' space rather than a gridded one\footnote{With respect to numerical precision}. A strategy of particles moving towards a global optimum is much faster than an evolutionary grid search. Therefore, MO-PSO is orders of magnitudes faster than MOEA/D in searching the Pareto-optimal solution closest to the ideal, and utilizes a significantly smaller number of evaluations of the single-objective minimization problem \eqref{prob:mop_moead}. This benefit is underlined by two specific updates to the imaging algorithm: We use a different evolutionary strategy (swarm intelligence instead of genetic operations), and solve the single-objective minimization problems \eqref{prob:mop_moead} by fast, gradient-based convex optimization algorithms rather the slower genetic evolution. In consequence, a single MO-PSO run in \textbf{a dataset with similar number of data points, corruptions, and coverage as the EHT 2017 data} can be finished on time-scales of minutes on a medium-level notebook.

\subsection{Single-objective}
Standard RML methods focus on finding the best image by balancing regularization terms and data terms with manually pre-defined parameter combinations. Hence, they need to explore a set of hyper-parameter combinations on synthetic data to find the best working combination. Then, they evaluate the final reconstruction with \textbf{these} parameter combinations. This exercise has been done intensively within the EHT data analyses \citep{eht2019d, eht2021a, eht2022c}. The EHT did not only include realistic image \textbf{structures} as synthetic data, but surveyed a larger number of geometric models and found a common parameter combination that worked across models. This proved crucial in establishing trustable, bias-free images of the shadows of black holes.

However, this exploration is expensive and \textbf{time consuming}. Also, it is limited \textbf{in the parameter combinations that are explored}, and does not encode relative or absolute information about their contribution. MO-PSO particularly differs and advances in two important ways from such a strategy.

First, we replace the grid-search technique inherent to parameter surveys by an evolutionary strategy based on swarm intelligence. MO-PSO needs only a couple of iterations to arrive at a parameter combination that passes the top-set selection criterion defined in \citet{eht2019d}. It speeds up the data analysis significantly in this way. Moreover, while the parameter survey strategy was sufficient to establish a robust image in total intensity, the problem quickly becomes \textbf{intractable} once polarization and time-dynamics is included due to high number of parameter combinations\textbf{,} scaling exponentially with the number of data terms. MO-PSO presents a better scaling since the number of iterations and number of particles only loosely depends on the number data terms, but \textbf{more strongly} on the complexity of the Pareto front and the number of their respective clusters \citepalias[see the discussion of the clusters in the Pareto fronts in][]{Mueller2023, Mus2023b}.

Second, MO-PSO includes a strategy to find the best weighting combination self-consistently on the data itself and does not incorporate studies on synthetic data. This criterion, the weighting combination that minimizes the distance to the ideal point, also allows for the reconstruction of parameter combinations that are specific to each data set, and thus the classification of data sets by their respective regularizer response, see Sec. \ref{sec: reg_characterization}.

\section{Verification \textbf{on EHT data}}\label{sec:synthetic_data}\label{sec:M87}

To verify the \textbf{performance} of the algorithm in Stokes I, we show the reconstructions of M87* observed \textbf{on} April 11 during the 2017 EHT campaign~\citep[see][]{eht2019d}, following the analysis done in~\citetalias{Mueller2023}. In the public repository (see Sec.~\ref{sec:softwareAvailability}) we \textbf{provide} the code to reconstruct other geometric forms.

\begin{figure}
    \centering
    \includegraphics[scale=0.6]{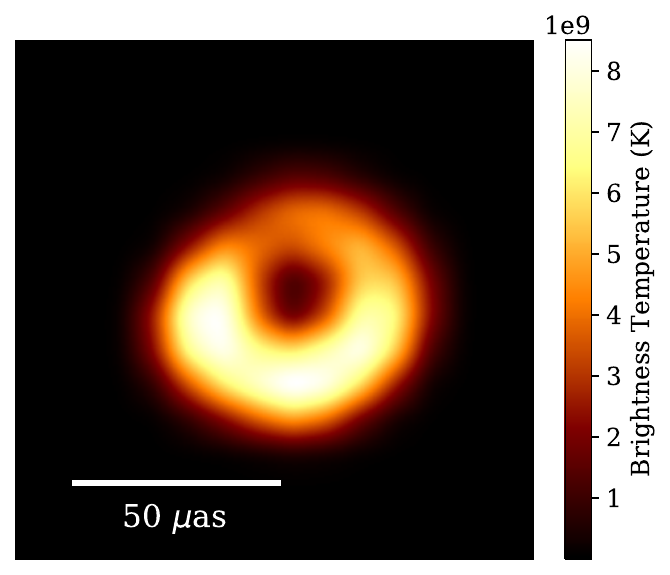}
    \caption{Reconstruction of M87 in April 11, 2017 using PSO algorithm, with 25 particles.}
    \label{fig:m87_stokesI}
\end{figure}

Figure~\ref{fig:m87_stokesI} shows the reconstructed M87* PSO image. We have compared the reconstructed image with the one presented to the public by the EHT~\citep{eht2019d} using the normal-cross correlation (or nxcorr)~\citep[see, for instance][]{Farah2022} \textbf{which} gives $\sim0.94$. The algorithm was run using 50 iterations and 25 particles, and took, in a computer of 12 cores, \textbf{i7 10th generation} $\sim10\,$min.

The marginal contribution of the weights \textbf{is}: $\zeta \sim 13.9\%, \theta\sim6\%, \tau\sim11.02\%, \eta\sim7.57\%, \epsilon\sim33.17\%, \kappa\sim16\%$. From these values, we can extract that the \textbf{flux regularizer, l1} are crucial to obtain a good image. 

In Sect.~\ref{sec: reg_characterization} we perform a statistical analysis of the marginal contribution of each weight in this M87* real data and in synthetic geometric models. We aim to generalize the conclusion here explained for each regularizer and constraint their range.

We will discuss and present the performance of MO-PSO in recovering polarimetric signatures as part of the dynamic and polarimetric movie reconstruction that is discussed in the next section. However, as an additional demonstration case, we redo the geometric ring model tests that were performed in \citetalias{Mus2023b} in Appendix \ref{app: polarimetry}.

\begin{figure}
    \centering
    \includegraphics[scale=0.7]{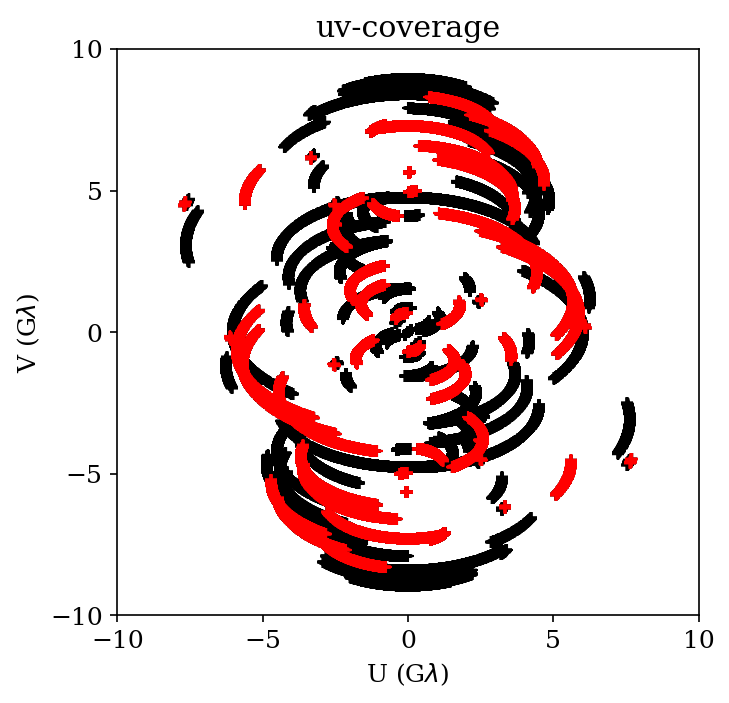}
    \caption{uv-coverage for SgrA* April 11 2017 EHT (red dots) and ngEHT at 230\,GHz (black crosses).}
    \label{fig:uv_cov}
\end{figure}

\section{Verification on synthetic data in EHT+ngEHT array}\label{sec:synthetic_data}

We test the capability to do dynamic polarimetry with a synthetic data set that is based on the third ngEHT Analysis challenges \citep{ngehtchallenge}\footnote{\url{https://challenge.ngeht.org/challenge3/}}.
The ground truth movie of Sgr A* is based on a model presented \citep{Broderick2016} with a shearing hotspot \citep{Tiede2020b} inspired by the observations of \citet{Gravity2018}. \textbf{The respective uv-coverage for the whole observation is presented in Fig. \ref{fig:uv_cov} (black dots). We add thermal noise, but assume a proper gain and leakage calibration. Note that since $\alpha=\beta=0$ is set as a configuration for this manuscript, compare Sec. \ref{sec: functionals}, the reconstruction in total intensity does not depend on the gains. Residual gain and leakage corruptions may affect the reconstruction in linear polarization. However, we have evaluated the impact of these corruptions in \citetalias{Mus2023b} for the same synthetic movie. We found that the effect of residual leakage errors is small, in line with the claims made by \citep{Marti2008}, marking the current data set sufficient for testing purposes. Finally, we would like to refer to ongoing work in recovering the polarimetric signal directly from the closure traces, a closure product that is independent against any gain and leakage corruptions \citep{Broderick2020b}. A first pioneering work has been done by \citet{Albentosa2023} as a modeling approach that has been extended to imaging with MOEA/D and MO-PSO in \citet{Mueller2024b}.}

\begin{figure*}
    \centering
    \includegraphics[width=\textwidth]{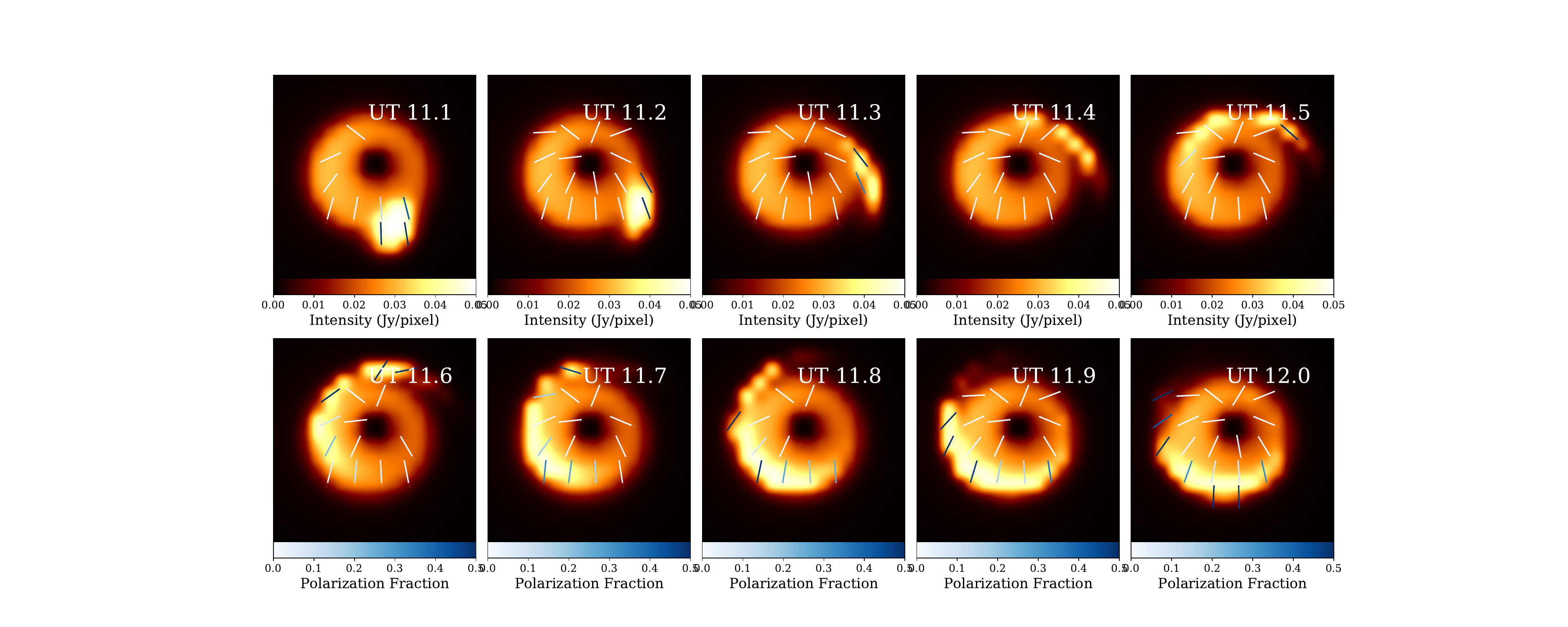}
    \caption{True movie of the ngEHT Challenge 3 synthetic data~\citep{ngehtchallenge}. Observation has been divided in 10 keyframe. EVPA are represented as lines and their corresponding polarization fraction appears as color map: the bluer, the stronger $m$. The intensity of the source is represented by other color map.}
    \label{fig: true_movie}
\end{figure*}

For more details on the simulation we refer to \citet{ngehtchallenge, Chatterjee2023}. We show the ground truth movie in Fig. \ref{fig: true_movie}.
\textbf{A synthetic observation of the model movie with the EHT+ngEHT array was conducted following the procedure outlined in }
\citet{ngehtchallenge}. This configuration consists of all current EHT antennas and ten additional antennas from \textbf{a} list of proposed sites by \citet{Raymond2021}.
We \textbf{split the observation into scans with 10 minutes in length and gaps of two minutes between scans}, as was done already in \citetalias{Mus2023b}.

\begin{figure*}
    \centering
    \includegraphics[width=\textwidth]{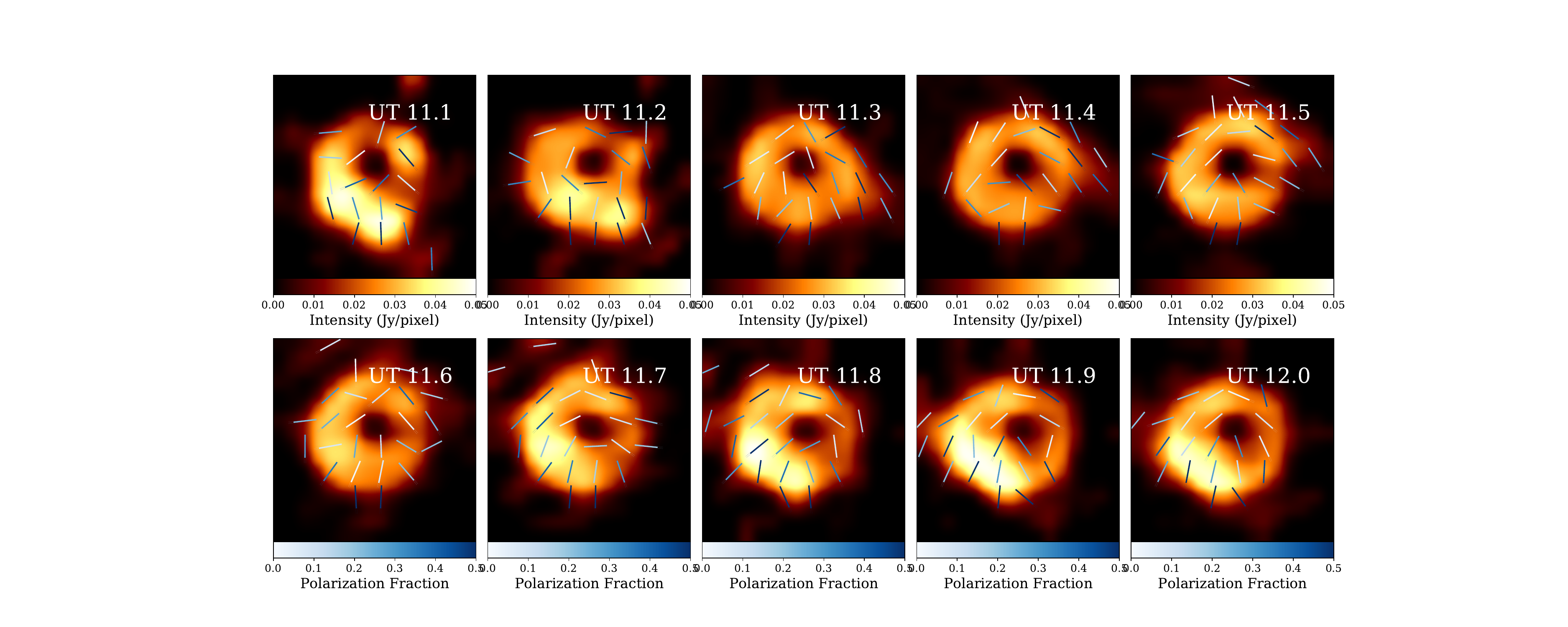}
    \caption{Recovered PSO movie of the ngEHT Challenge 3 synthetic data~\citep{ngehtchallenge}.}
    \label{fig: reco_movie}
\end{figure*}

\begin{figure}
    \begin{subfigure}{0.5\textwidth}
    \hspace{-2cm}
    \includegraphics[scale=0.45]{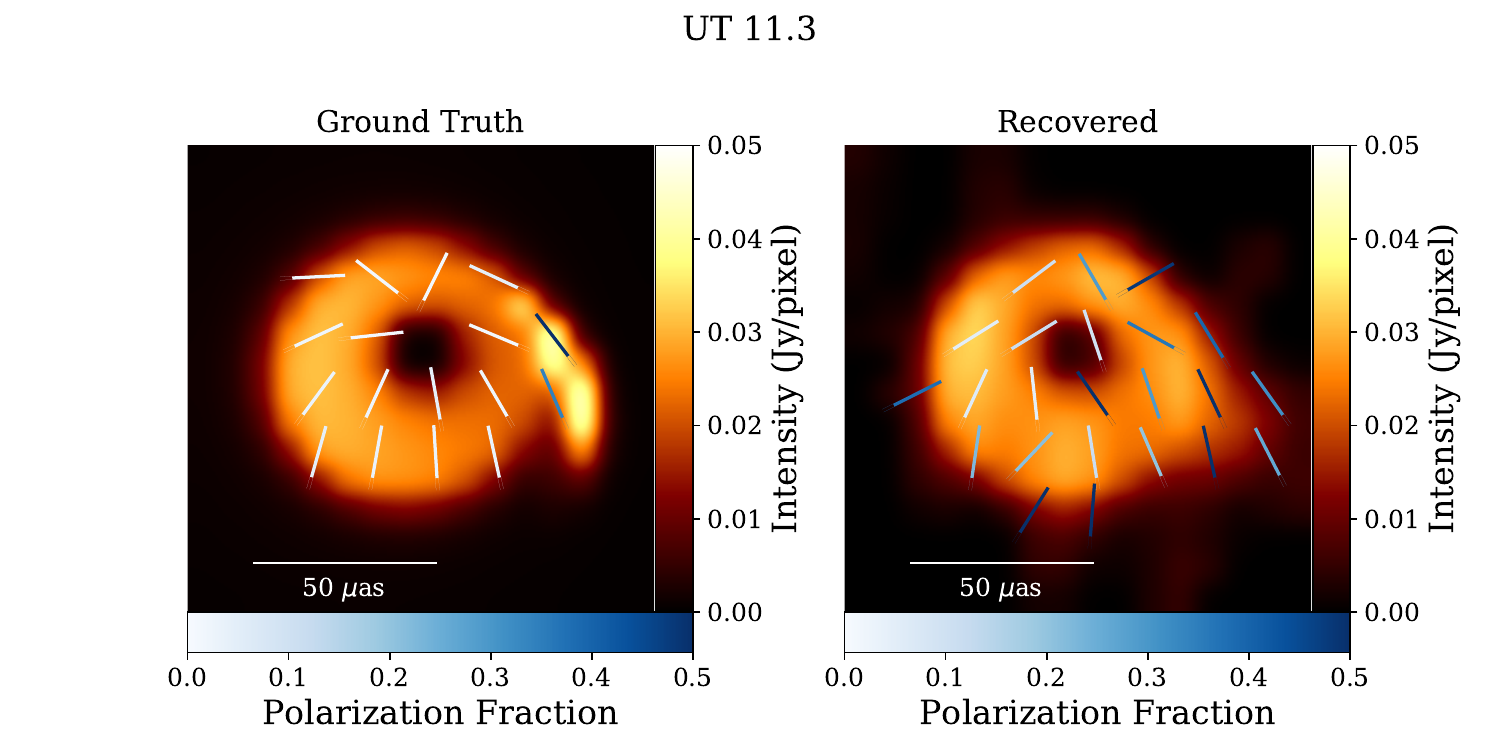}
    \end{subfigure}
    \begin{subfigure}{0.5\textwidth}
    \hspace{-2cm}
    \includegraphics[scale=0.45]{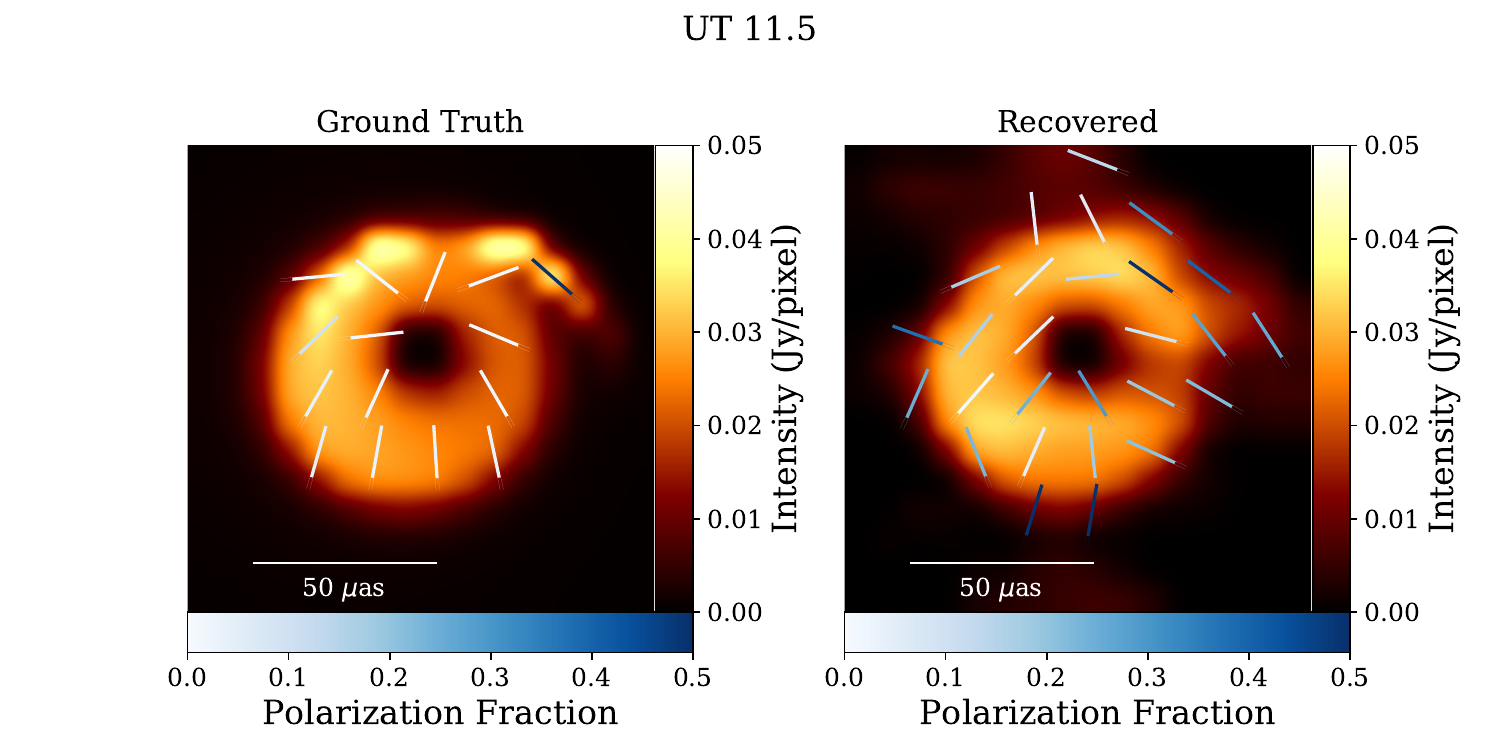}
    \end{subfigure}
    \begin{subfigure}{0.5\textwidth}
    \hspace{-2cm}
    \includegraphics[scale=0.45]{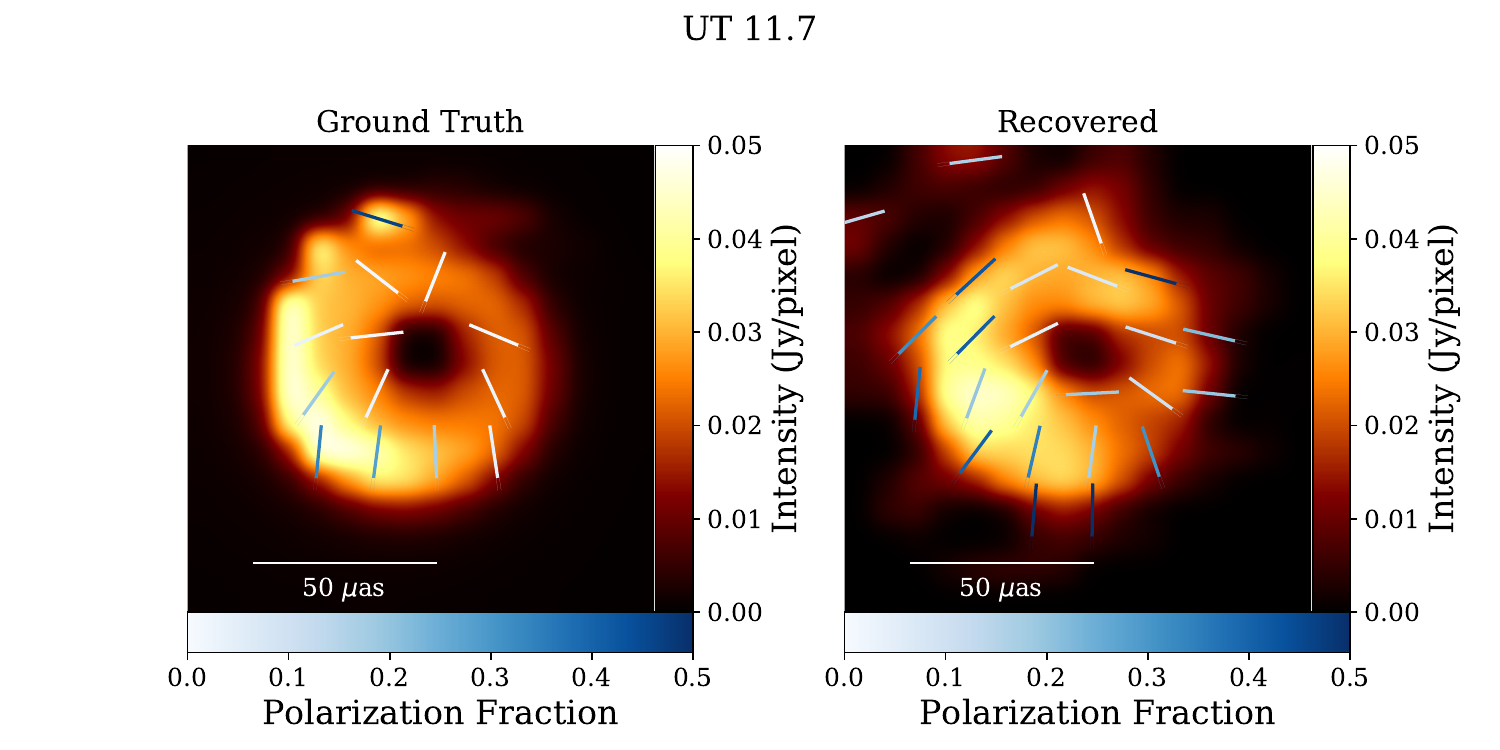}
    \end{subfigure}
    \caption{Keyframe comparison between the true movie (left panel)  and recovered PSO movie (right panel). Keyframes times are on the top of each pair.}
    \label{fig: compplots}
\end{figure}

We show the recovered reconstruction in Fig. \ref{fig: reco_movie}. The polarimetric movie was recovered with snapshots of six minutes in the time-window with the best uv-coverage \textbf{as reported by \citet{ngehtchallenge}. The selection of the time window was based on the coverage analysis metrics presented in \citet{Farah2022}.}. For better comparison we show single snapshots in Fig.~\ref{fig: compplots}. The ring-like image of the black hole shadow is successfully recovered at every keyframe with varying brightness asymmetry. Moreover, we successfully recover some hints for the dynamics at the event horizon in the form of a shearing hotspot. The overall EVPA pattern (orientation and strengths) and \textbf{its} dynamics are \textbf{successfully} recovered in all keyframes. \textbf{There are however minor differences in the plotted EVPA ticks. Here we would like to note that the ticks at the edge of the ring a supported in a region with limited total intensity and are less reliable. Furthermore, the groundtruth model contains structures on smaller scales than the beam resolution, which makes the recovered polarization structure appearing more scrambled. In Fig. \ref{fig: reco_movie_coarse}, we show the recovered and ground truth model when regridded to a smaller resolution, showing a significantly improved match between the synthetic data and the reconstruction}. 

We presented the same figures in \citetalias{Mus2023b} for the reconstruction with MOEA/D\footnote{Authors refer to~\citetalias{Mus2023b} for reconstructions of this same synthetic data-set using MOEA/D.}. However, note that MOEA/D is limited by the number of pixels as discussed in \citetalias{Mus2023b} due to issues with the numerical performance. These issues have been waived for MO-PSO as discussed in Sec. \ref{ssec: comp_to_moead}. Therefore, we recover the movie with a larger number of pixels, and effectively a better resolution than the reconstruction that was presented in \citetalias{Mus2023b}. However, to facilitate performance comparisions with the MOEA/D reconstruction, we present in Appendix \ref{app: coarse} the recovered movie regridded to the resolution of MOEA/D.

\begin{figure}
    \centering
    \includegraphics[width=0.5\textwidth]{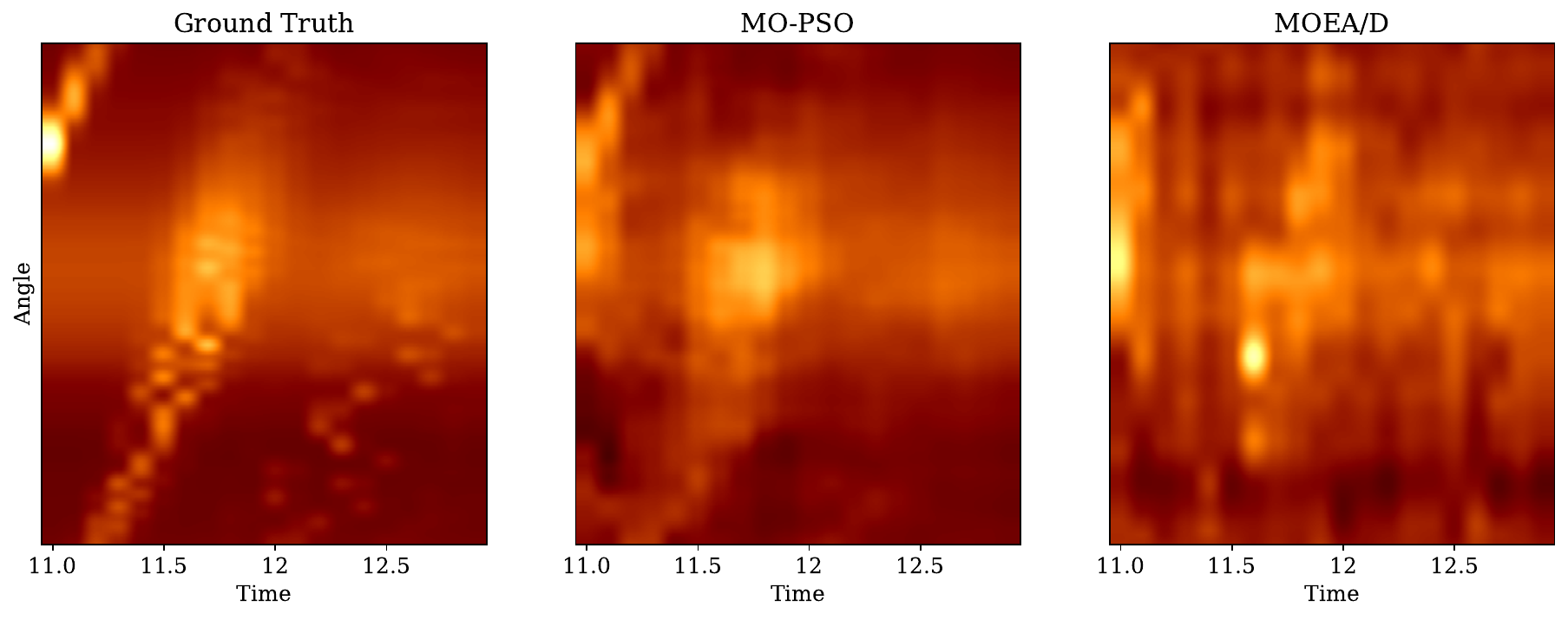}
    \caption{Phase-diagram that shows the angular brightness distribution as a function of time comparing the ground truth movie (left panel), and the recovered movie (right panel).}
    \label{fig:profiles}
\end{figure}

\begin{figure}
    \centering
    \includegraphics[width=0.5\textwidth]{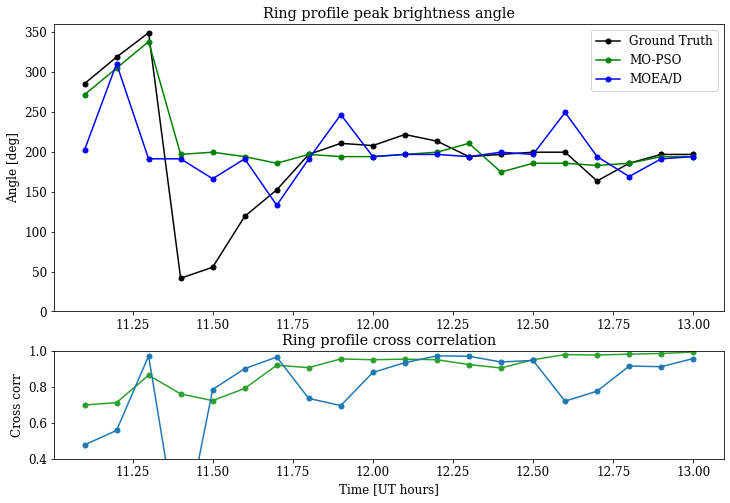}
    \caption{The top panel shows the angle of the brightest pixel in the ring as a function of time comparing the ground truth and the recovered solution. The bottom panel depicts the cross-correlation between the ground truth and the recovered solution in every scan.}
    \label{fig:pa}
\end{figure}

To quantify the accuracy of the reconstruction, and allow the comparison with previous works, we present here the same metrics that were used already in \citetalias{Mus2023b}. \textbf{In Fig. \ref{fig:profiles}} we represent the angular time evolution by the phase-diagram proposed by \citet{Mus2023}, \textbf{while in Fig. \ref{fig:pa}} we show the angle of the brightest pixel in the ring (tracking the hotspot movement) and the cross-correlation in every scan (in total intensity). These figures suggest that the overall angular distribution is recovered well in every scan except for some frames at the beginning \textbf{where} the reconstructions \textbf{lack} behind the rapid evolution of the ground truth. For comparisons, we also reprint the scoring of the MOEA/D reconstructions in Fig. \ref{fig:profiles} and Fig. \ref{fig:pa} from \citetalias{Mus2023b}. MOEA/D \textbf{recovered} the movie at a smaller resolution which explains some of the differences, particularly the coarser structure in the phase-diagram Fig. \ref{fig:profiles}. To allow for quantitative comparisons, we show the cross-correlation to the ground truth movie blurred to the MOEA/D resolution for MOEA/D in Fig. \ref{fig:pa}. \textbf{MOEA/D performs well in recovering the structure and angular evolution in most of the keyframes.}. The phase-diagram indicates that the reconstruction fidelity has been greatly improved by MO-PSO which is mirrored by the superior cross-correlation in almost every scan. Notably, the cross-correlation of the MO-PSO has relatively large values ($\sim 0.8$) \textbf{in} the first scans of rapid evolution \textbf{in} which MOEA/D failed. 

\begin{figure}
    \centering
    \includegraphics[width=0.5\textwidth]{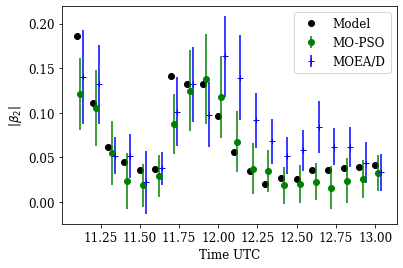}
    \caption{The amplitude of $\beta_2$ as a function of time comparing the ground truth movie (black) and the recovered movie (green). The errorbars represent the alignment error.}
    \label{fig:beta2_comp}
\end{figure}

\begin{figure}
    \centering
    \includegraphics[width=0.5\textwidth]{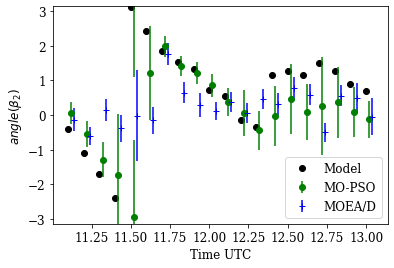}
    \caption{The phase of $\beta_2$ as a function of time. The errorbars represent the alignment error.}
    \label{fig:beta2_angle}
\end{figure}

\begin{figure}
    \centering
    \includegraphics[width=0.5\textwidth]{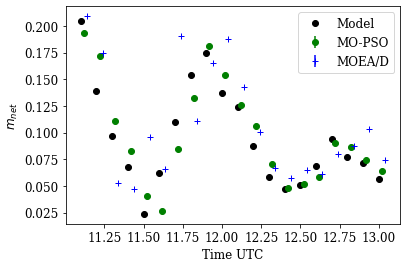}
    \caption{The net polarization as a function of time.}
    \label{fig:mnet}
\end{figure}

The quality of the polarimetric reconstructions as a function of time, are shown in Fig. \ref{fig:beta2_comp}, Fig. \ref{fig:beta2_angle} and Fig. \ref{fig:mnet}. Fig. \ref{fig:beta2_comp} and Fig. \ref{fig:beta2_angle} depict the amplitude and phase of the $\beta_2$ parameter introduced by \citet{Palumbo2020}. $\beta_2$ traces the orientation, twistiness and amplitude of the polarimetric signal. Additionally, Fig. \ref{fig:mnet} illustrates the net polarization as introduced in Sect. 5.3~\citet{eht2021a}. The polarized feature \textbf{are} recovered very well, even for the rapidly evolving first part of the movie. \textbf{We show the reconstructions of the polarimetric quantities by MOEA/D from \citetalias{Mus2023b}, and note that MO-PSO performs better}. However, note that there is the effect of a higher resolution \textbf{is} visible again, i.e. we compare the MOEA/D reconstructions against a ground truth movie at the higher MO-PSO resolution. Since the computation of $\beta_2$ includes an alignment of the images on the same center, there is a slight error that could occur from the alignment of the reconstruction to the \textbf{ground} truth. We estimated the error of the alignment by misaligning the images by one pixel towards the North, East, West and South and recomputing $\beta_2$ applying the same strategy that was adapted in \citetalias{Mus2023b}. The true $\beta_2$ and recovered $\beta_2$ match in almost all scans within one standard deviation.
 
\textbf{We achieve a successful reconstruction of a dynamic, polarimetric movie with a time-regularization approach that is simpler than some other works\citep[e.g.][]{Bouman2017, Leong2023}}. However, as already demonstrated in \citetalias{Mus2023b} MO-PSO does not need sophisticated prior information \textbf{for} temporal regularization, it rather maps the uncertainties in the scan-to-scan variability in the Pareto front (instead of the regularizer) and navigates the temporal correlation by investigation of the Pareto front.

\section{Regularizers characterization}\label{sec: reg_characterization}

MO-PSO presents two main advantages in comparison with MOEA/D:

On the one hand, there is no need to grid the hyper-parameter space (we let the weights \textbf{change} during the optimization process) and, with a high number of pixels, this discretization makes the \textbf{resulting} image \textbf{artificially pixelated}. When a thin mesh is used, each pixel has its own weight combination. Moreover, the higher the number of pixels or the number of objectives, the more complex ~\citep{curry2014} the whole MOEA strategy is. In this algorithm, the number of pixels only affects in the optimization step.

On the other hand, to study each regularizer separately \textbf{allows us to investigate} the marginal contribution of each \textbf{regularizer} in the final image. In other words, we can know how important are each $f_i$ and how they \textbf{affect} the final geometry. \textbf{However, in contrast to full Bayesian exploration techniques, the final distribution of the particle swarm should not be interpreted as a posterior in the sense of Bayesian imaging.}

We have executed Algorithm~\ref{alg:shapso} on multiple instances, initiating it 100 times with 25 new particles in each run (utilizing the randomized nature of particle swarm optimization with multiple seeds). From these iterations, we have saved the optimal weight values associated with the algorithm's convergence. Our aim was to examine the distribution of these optimal weights to \textbf{explore} their variance and its correlation with the quality of the resulting images.

Our anticipation was that if the images obtained were similar in quality, the respective optimal weights would exhibit close proximity. A notable standard deviation among these weights, while still yielding comparable reconstructed images, suggests that specific regularizers might not significantly impact the reconstruction. This observation stems from the notion that despite variations in their weights, the resulting images maintain \textbf{consistency}.

\subsection{Synthetic EHT-like data}

\begin{figure*}
    \centering
    \hspace{-1.5cm}
    \includegraphics[scale=.32]{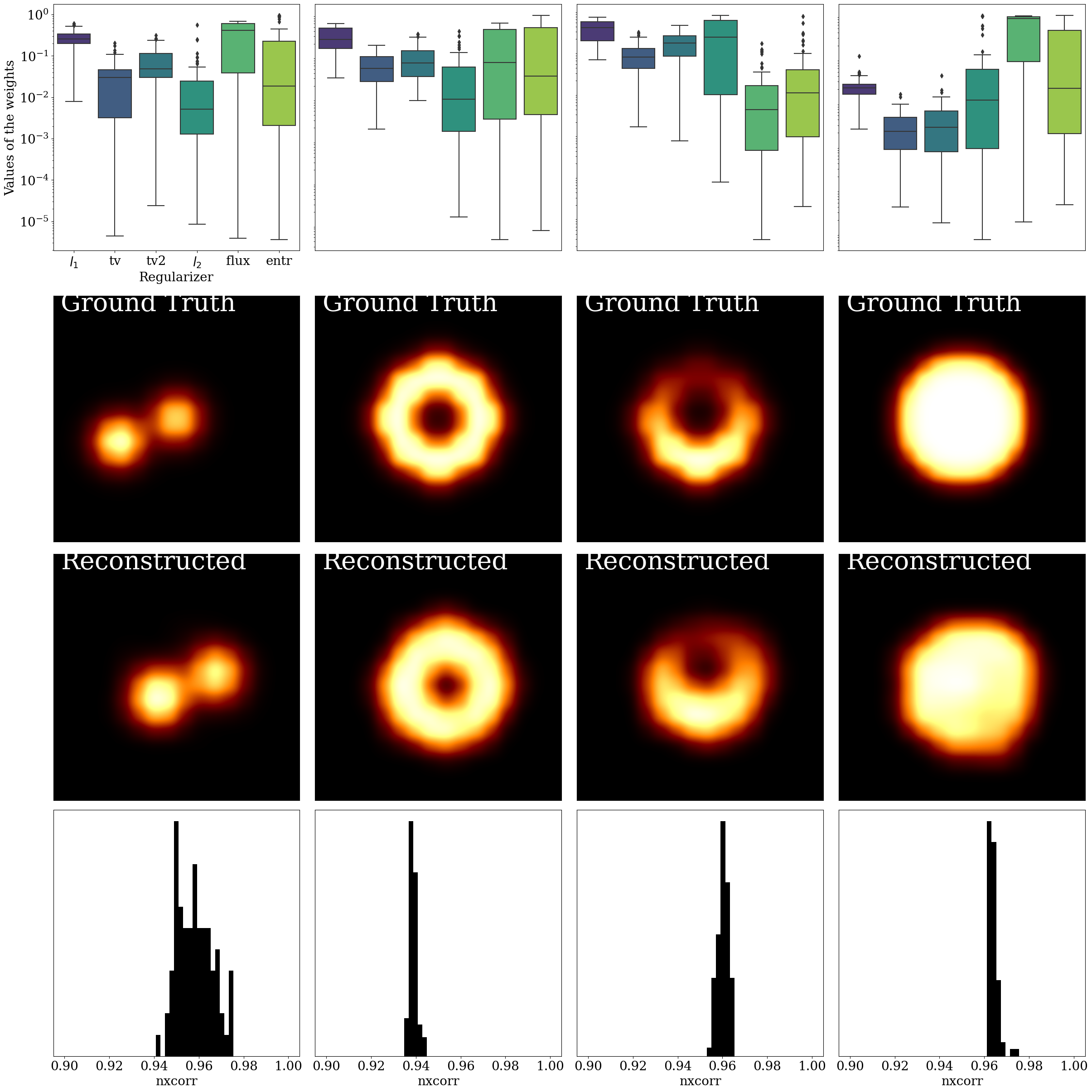}
    \caption{\textit{Top row:} Marginal contribution of each regularizer for each model (double, ring, crescent and disk - from left to right). Box represent the quartiles, black line the mean, whiskers the rest of the distribution. Empty circles are outliers. \textit{Second and third row:} Ground truth and the (best, in terms of nxcorr) reconstructed image. \textit{Bottom row:} nxcorr distribution for every element of the survey.}
    \label{fig:all_regularizers}
\end{figure*}

We have conducted a survey involving the reconstruction of four standard geometric models (double, ring, crescent, and disk) w\textbf{ith synthetic observations using the 2017 April 11 EHT coverage towards \sgra.} The results are presented in Fig.~\ref{fig:all_regularizers}.

The top row (R1) displays box-plot distributions of \textbf{the} marginal distributions of each regularizer: $l_1$-norm, tv, tsv, $l_2$-norm, flux, and entropy, for each model. The $l_1$-norm generally is in the interval $\left(0.1,1\right)$ for 75\% of cases, signifying its higher impact compared to entropy, which exhibits a wider range (being less critical in the final image reconstruction). Notably, in the case of the ring model, the flux regularizer takes precedence over others, playing a pivotal role in achieving a high-quality image.

\textbf{The} second and third row show the model images and the best reconstructed image, in terms of the nxcorr for each case.

\textbf{The} last row (R4) presents the nxcorr distribution for every model. In all cases, \textbf{all the weight combinations in the final population of the swarm showcase a nxcorr exceeding 0.92}. Although the double model exhibits a slightly broader distribution with a mean nxcorr around 0.94, the remaining images display minimal nxcorr dispersion, indicating their satisfactory quality in each run of MO-PSO. This also demonstrates the performance and reconstruction accuracy of MO-PSO. Across multiple geometric models, and in every single instance of the particle swarm optimization, MO-PSO finds a suitable reconstruction of the image, and a parameter combination that would pass the top-set selection criterion for this single source defined by \citet{eht2019d}.

\subsection{M87 real data}\label{subsec:M87_marginal}

\begin{figure}
    \centering
    \includegraphics[scale=.15]{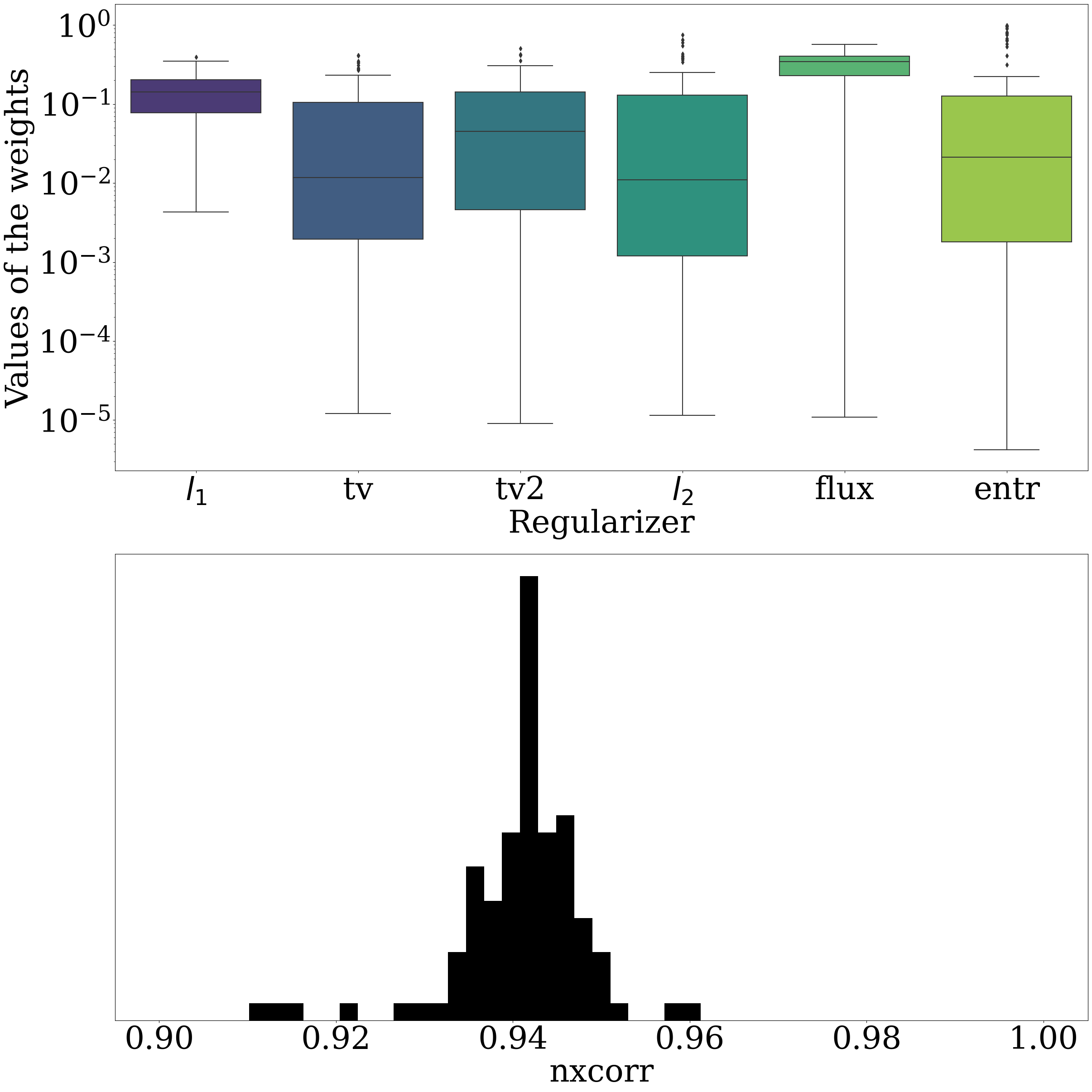}
    \caption{\textbf{Top panel: }Marginal contribution of each of the regularizers for the M87 April 11 image. Black continuous line represents the mean, and the boxes contain the 75\% of the values. \textbf{Bottom panel: nxcorr distribution. Most of images have nxcorr $\sim 0.94$. The nxcorr of the images ranges between $\sim0.91$ and $\sim0.96$.}}
    \label{fig:M87_regularizers}
\end{figure}

\textbf{We show in} Fig.~\ref{fig:M87_regularizers} the relative importance of the regularizers for M87* (see Sect,~\ref{sec:M87}).

We first note that all reconstructions have nxcorr with respect to the published image by the EHT~\citep{eht2019d} greater than 0.9 and a mean of $~\sim 0.94$. This serves as a convincing alternative verification of the EHT results on M87. The intrinsic morphology of the source makes its reconstruction a bit more challenging than the synthetic geometric models, as expected. 

Second, we can observe that the weights associated \textbf{with} the $R_{flux}$ and $R_{l_{1}}$ are more constrained, but the rest of hyper-parameters have less dispersion than the geometric models while the entropy and $l^2$ accept greater variation. Indeed, the range of better values for the regularizers is $\sim\left(1e^{-3}, 0.5\right)$. The box-plot representation is more similar to the double, ring, or crescent, than to the disk. Regularizers used for the double, ring or crescent are better option than the ones for the disk.

\section{Summary and Conclusions}
\label{sec:summary}

Imaging reconstruction of interferometric data is a challenging ill-posed problem, particularly when dealing with very sparse uv-coverage, as is often the case in global VLBI observations. These challenges are amplified when working with polarimetric data or attempting to capture the dynamic behavior of the source, \textbf{and become even more complex when attempting to reconstruct all these aspects together}.

While static polarimetry imaging has been extensively studied in the past~\cite{Ponsonby1973,Narayan1986,Holdaway1988,Coughlan2013,Chael2016}, solving the non-convex problem of polarimetric \textbf{dynamic} imaging remains an open challenge.

The intricacies presented by rapidly evolving sources, such as the case of \sgra~\citep{eht2022c} have motivated the VLBI community to develop innovative algorithms capable of effectively addressing variability concerns. The inherent limitations of snapshot imaging due to restricted coverage and the potential loss of information resulting from temporal averaging highlight the necessity for the formulation of functionals that can efficiently mitigate such information loss.

In previous works,~\citetalias{Mueller2023,Mus2023b} introduced the MOEA/D algorithm, a full Stokes dynamic multiobjective optimization problem to recover unsupervised a set of local minima. However, this algorithm has some limitations. In particular, its high computational complexity and the gridding of the space could provoke different pixels in the same image to have different associated weights, \textbf{producing} reconstructions with artifacts. Moreover, an \textit{ad-hoc} rescaling value \textbf{was needed} to avoid numerical problems during the genetic operations.

The alternative, classical RML approaches paired with a parameter survey strategy proved successful in robustly \textbf{recovering} a wide range of image features even for very challenging data sets. They are fast \textbf{at} producing a single maximum-a-posteori solution given a specific set of weights, but the survey strategy takes time and relies on synthetic data.

In this work, we have developed a novel approach, MO-PSO, to solve the \textbf{image reconstruction} problem aiming to combine the best of RML and MOEA/D. MO-PSO keeps most of the advantages of MOEA/D as a multiobjective, easy adaptable, global and self-consistently unsupervised search technique and overcomes its limitations with regard to numerical complexity and gridding. We seek the optimal weights of a multi-objective optimization problem, and then we obtain \textbf{the} associated image. One main advantage of this approach is that finding the optimal weights is a convex problem, and thus, it has a unique global solution.

In addition, our algorithm is as flexible as MOEA/D, and allows to study and to characterize the marginal contribution of each regularization term with respect to the final (full Stokes dynamic/static) reconstruction.

\begin{acknowledgement}
HM and AM have contributed equally to this work. This work was partially supported by the M2FINDERS project funded by the European Research Council (ERC) under the European Union’s Horizon 2020 Research and Innovation Programme (Grant Agreement No. 101018682) and by the MICINN Research Project PID2019-108995GB-C22. AM also thanks the Generalitat Valenciana for funding, in the frame of the GenT Project CIDEGENT/2018/021. HM received financial support for this research from the International Max Planck Research School (IMPRS) for Astronomy and Astrophysics at the Universities of Bonn and Cologne.\\
\\
\textbf{Software Availability}\label{sec:softwareAvailability}
Our imaging pipeline and our software is included in the second release of MrBeam \footnote{\url{https://github.com/hmuellergoe/mrbeam}}. Our software makes use of the publicly available ehtim \citep{Chael2016, Chael2018}, regpy \citep{regpy}, MrBeam \citep{Mueller2022, Mueller2022b, Mueller2022c} and pyswarms \footnote{\url{https://pyswarms.readthedocs.io/en/latest/}} packages.
\end{acknowledgement}

\bibliographystyle{aa}
\bibliography{lib}{}

\appendix
\section{Polarimetry} \label{app: polarimetry}

To demonstrate the capabilities of PSO imaging in the polarimetric domain, we replicate the experiments presented in \citet{Mueller2022c} and~\cite{Mus2023b}, using a synthetic data source \textbf{from} the \textit{ehtim} software package \citep{Chael2016, Chael2018}. 

For the polarimetric reconstruction, we follow \textbf{a similar strategy that we used for~\citetalias{Mus2023b}, where} we fixed the Stokes I reconstruction, and only solved for linear polarization. We initialized the initial population with images with constant electric vector position angle (EVPA) at a constant polarization fraction of $1\%$ across the whole field of view. Rather than \textbf{solve for the $Q$ and $U$ separately,} we equivalently model the linear polarization by the linear polarization fraction $m$ and the mixing angle $\chi$, i.e.:
\begin{align}
    Q + iU = m \cdot I \cdot e^{2\pi i \chi}.
\end{align}
This turns out to be beneficial since inequality \eqref{eq: stokes_inequality} is automatically satisfied by imposing $0 \leq m \leq 1$, specifically we cannot create spurious polarimetric signals outside of the total intensity contours.

\begin{figure*}
    \centering
    \includegraphics[scale=0.5]{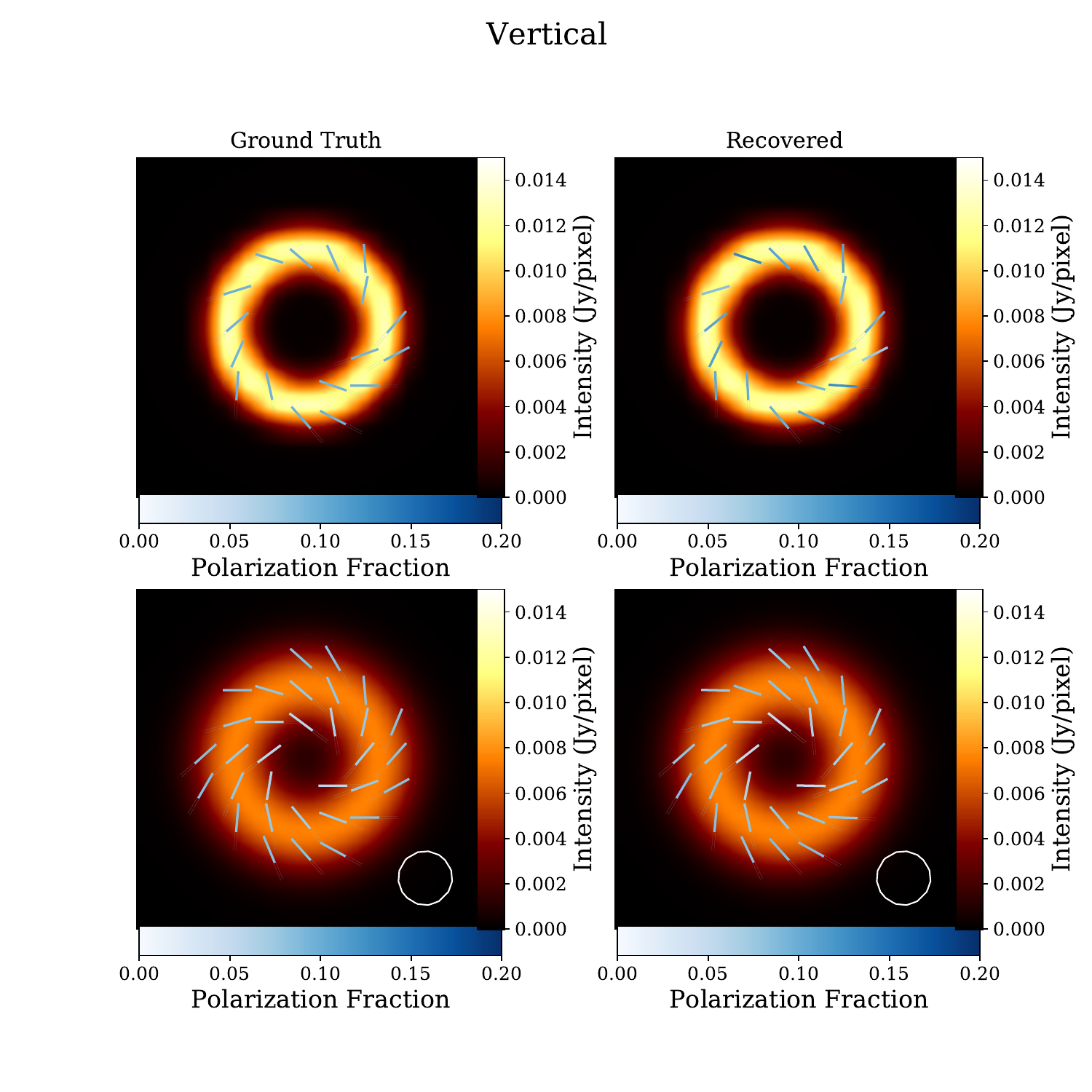}
    \caption{Simulated ring-like source with vertical EVPA using \textbf{the} EHT 2017 \textbf{coverage and} array \textbf{properties}. Left panel: Ground truth. Right panel: PSO reconstrucion (solution). The top panel show the reconstructions at full resolution, the bottom panels blurred with a resolution of $20\,\mu\mathrm{as}$. Vertical lines represent the EVPA and \textbf{the lone color represents} the amount of polarization fraction. The color map for the brightness of the source is different to the one of the EVPAs.}
    \label{fig: rings1}
\end{figure*}

\begin{figure*}
    \centering
    \includegraphics[scale=0.5]{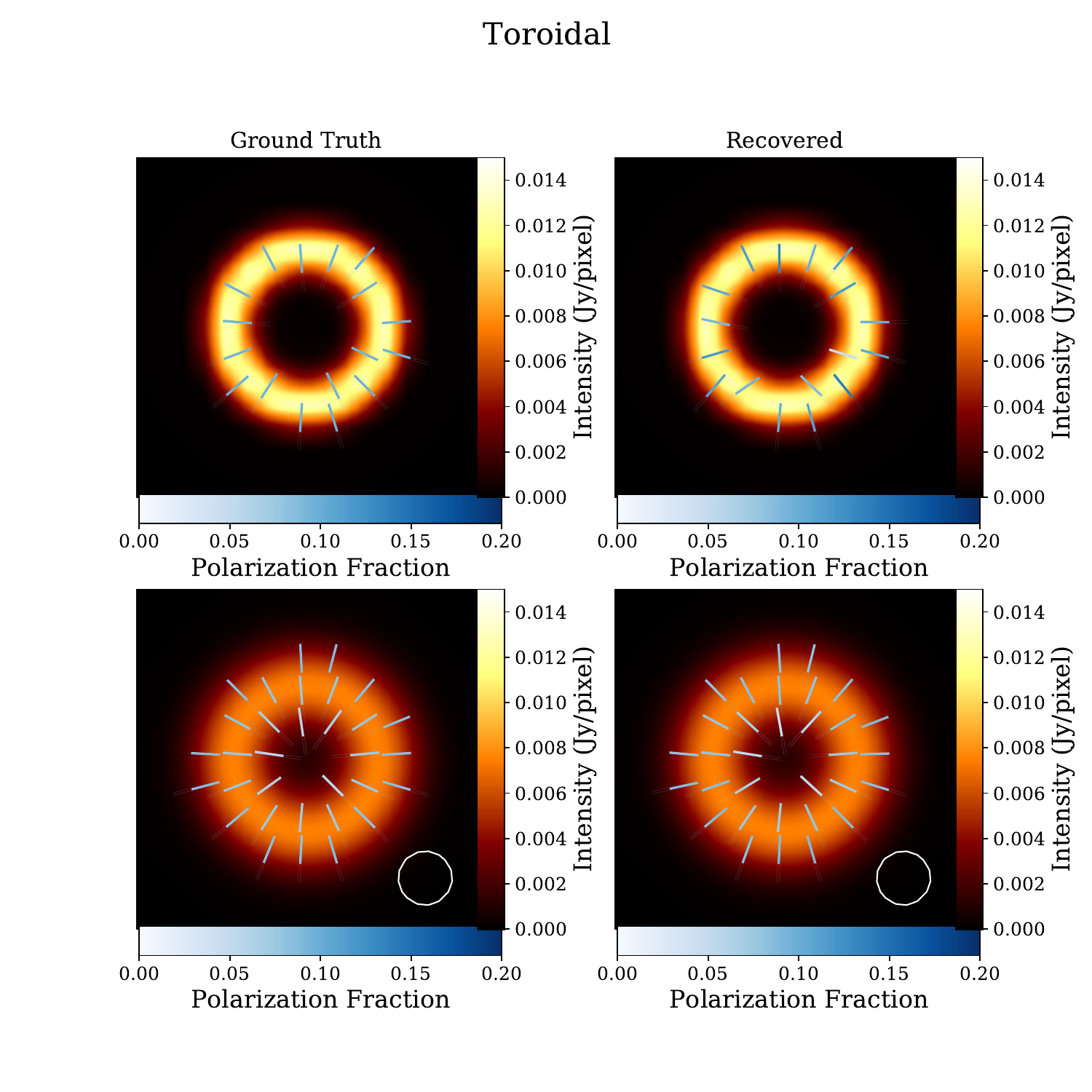}
    \caption{Same as Fig. \ref{fig: rings2}, but for a toroidal magnetic field.}
    \label{fig: rings2}
\end{figure*}

\begin{figure*}
    \centering
    \includegraphics[scale=0.5]{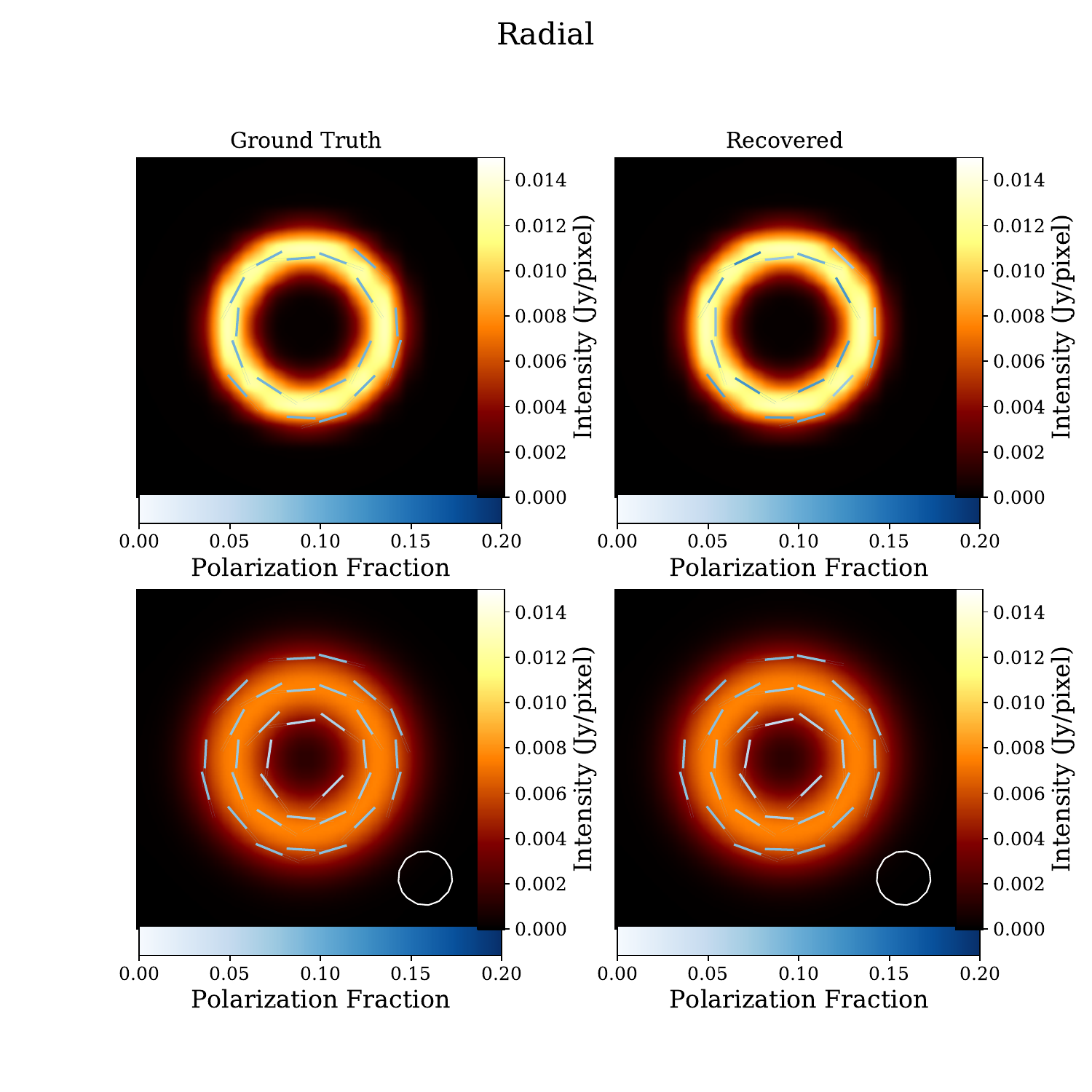}
    \caption{Same as Fig. \ref{fig: rings2}, but for a \textbf{radial} magnetic field.}
    \label{fig: rings3}
\end{figure*}

To verify that PSO reconstructions performs well for a wider range of EVPA structures, we performed a number of additional tests with an artificial ring image and different EVPA patterns (\textbf{vertical} magnetic field, \textbf{toroidal} magnetic field and a \textbf{radial} magnetic \textbf{field}). Fig.~\ref{fig: rings1}, Fig. \ref{fig: rings2} and Fig. \ref{fig: rings3} show the true model (left panel) and recover PSO image (right panel). The reconstruction of the overall pattern is successful in all four cases. In conclusion, we can differentiate various magnetic field configurations by the PSO reconstructions of EHT-like data.

\section{Recovered Movie at MOEA/D resolution} \label{app: coarse}
\begin{figure*}
    \centering
    \includegraphics[width=\textwidth]{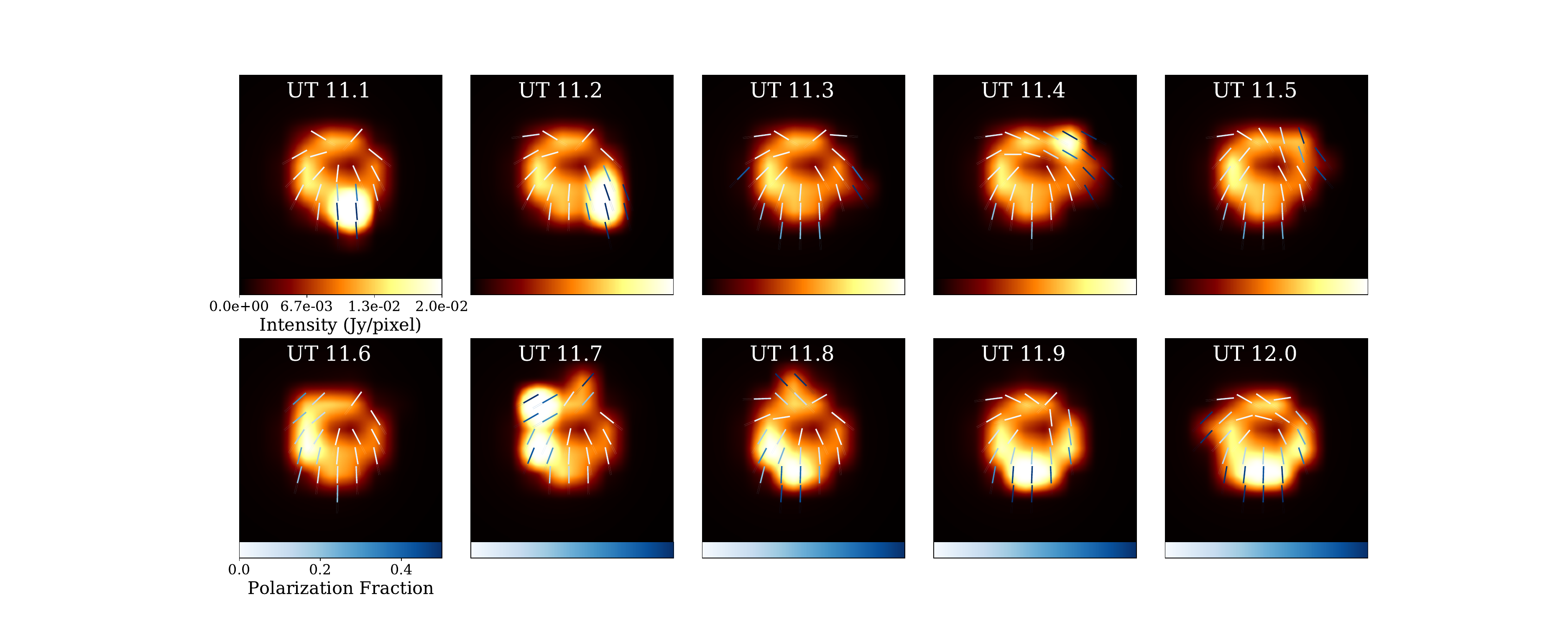}
    \includegraphics[width=\textwidth]{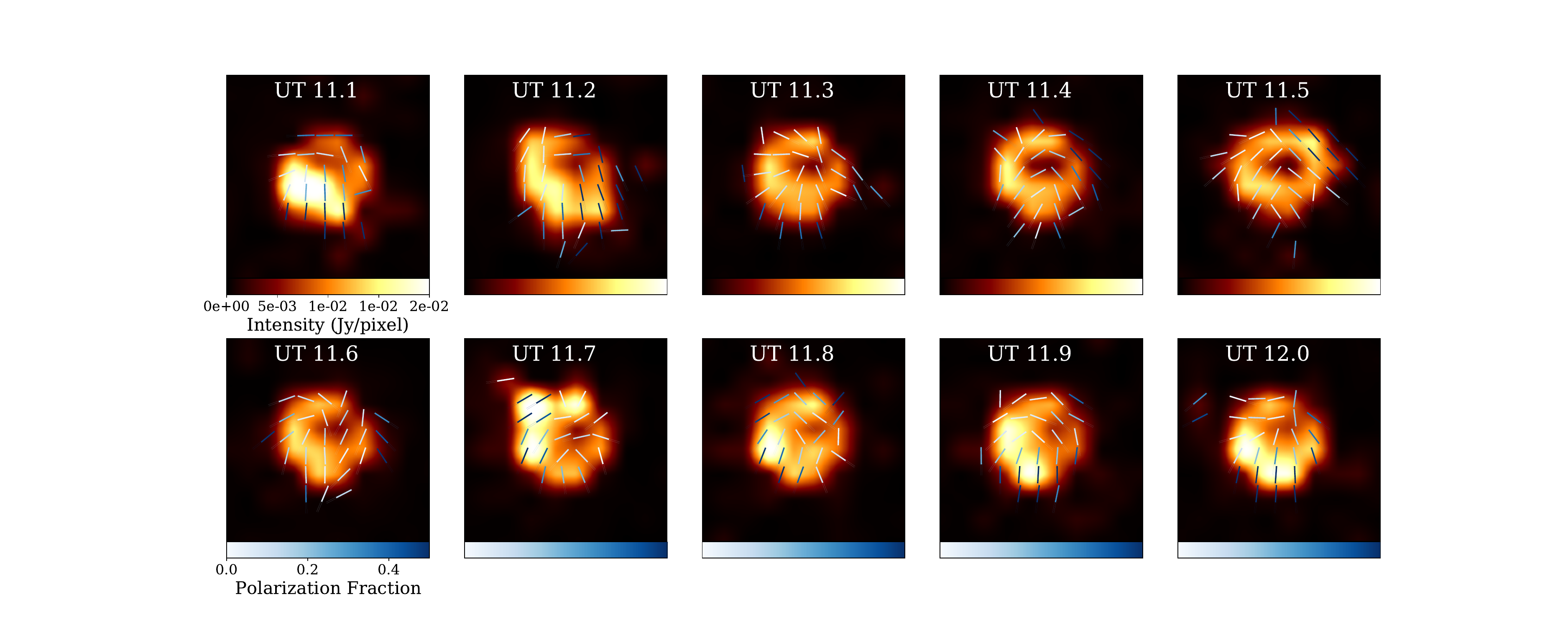}
    \caption{Same as Fig. \ref{fig: reco_movie}, but regridded to the field of view $180\mu\mathrm{as}$ and resolution $10\mu\mathrm{as}$ as was used in \citetalias{Mus2023b}. \textbf{Top: the true move. Bottom: the recovered.}}
    \label{fig: reco_movie_coarse}
\end{figure*}

We show the recovered movie with MO-PSO in Fig. \ref{fig: reco_movie}. The same observation data have been recovered already in \citet{Mus2023b} with MOEA/D, although with a smaller resolution due to the limitations with respect to numerical performance. To simplify comparisons we present in Fig. \ref{fig: reco_movie_coarse} the recovered movie with MO-PSO, but regridded to the same resolution and field of view as presented in \citetalias{Mus2023b}.

\subsection*{\textbf{Dynamics null test}}

\begin{figure*}
    \hspace{-2cm}
    \includegraphics[width=1.2\textwidth]{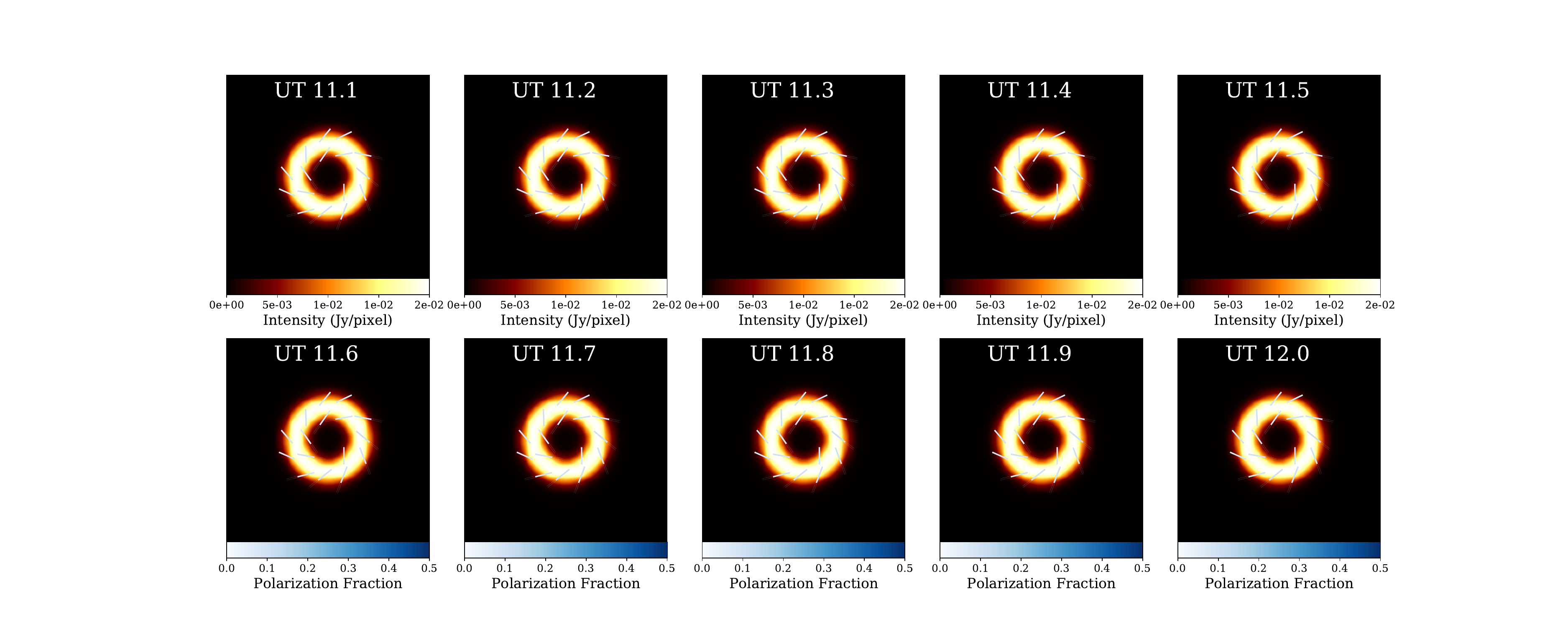}
    \caption{Static movie reconstruction using same uv-coverage and frames than in Fig.~\ref{fig: reco_movie}.}
    \label{fig:null_test}
\end{figure*}

We show how time regularizer is not creating an unexpected dynamic behavior. If the source is static, ngMEM is not biasing the image. To this end, we have simulated a ring using the ngEHT+EHT array displayed in Fig.~\ref{fig:null_test}, and we divided the observation in the same keyframes of Fig.~\ref{fig: reco_movie}.

\begin{figure}
    \centering
    \includegraphics[width=0.5\textwidth]{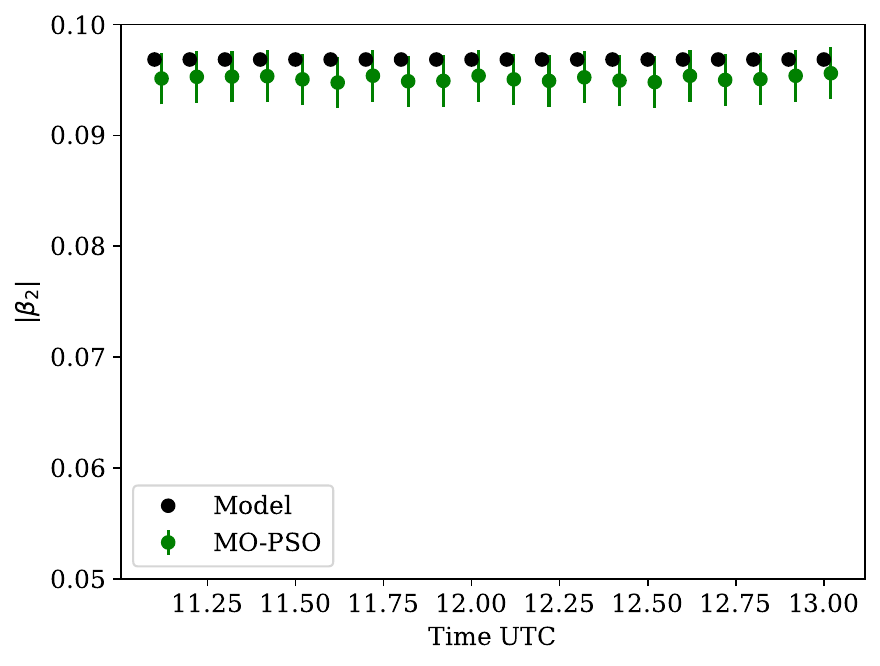}
    \includegraphics[width=0.5\textwidth]{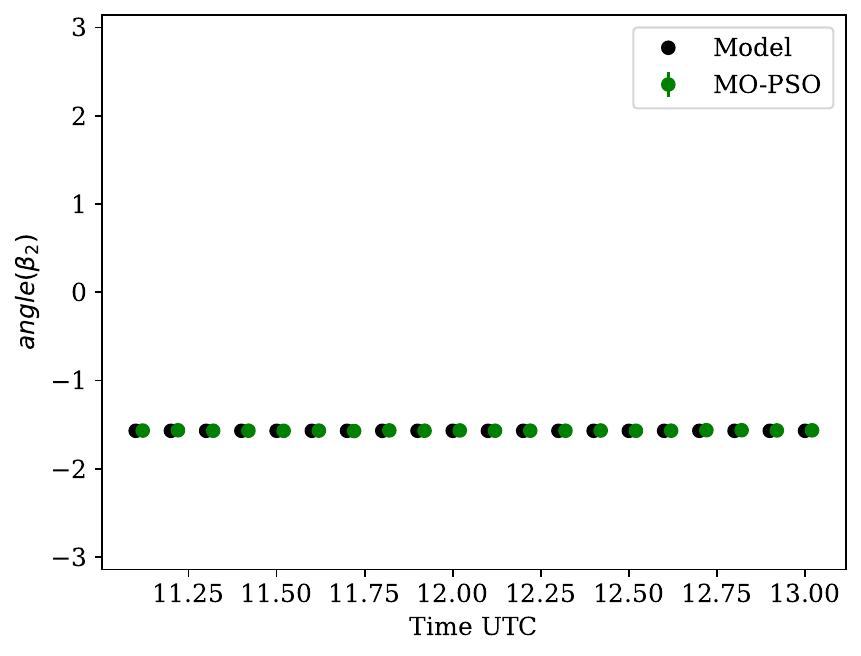}
    \includegraphics[width=0.5\textwidth]{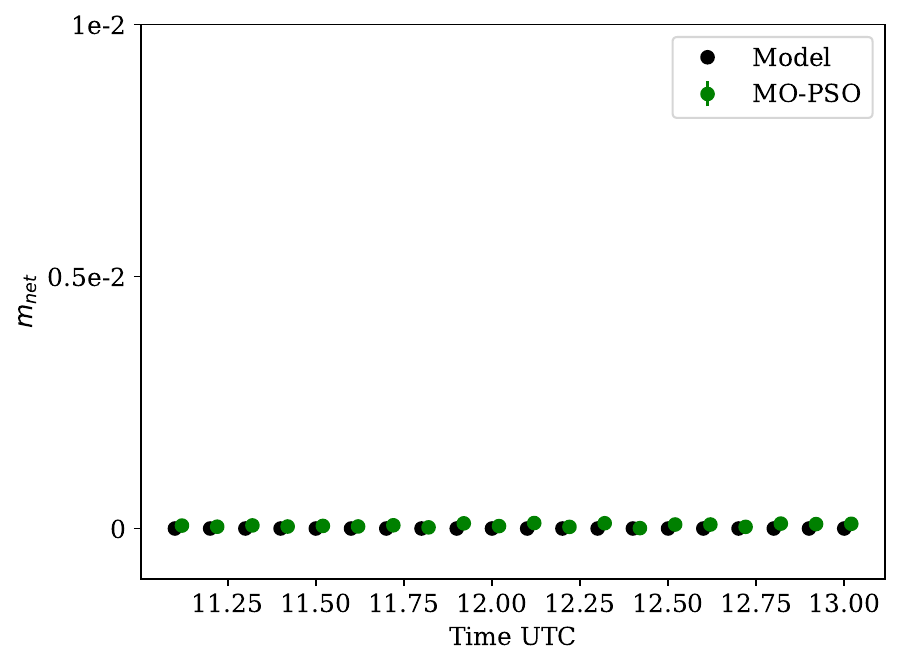}
    \caption{Same figures as Fig.~\ref{fig:beta2_comp},~\ref{fig:beta2_angle} and Fig.~\ref{fig:mnet}.}
    \label{fig:static_movie_stats}
\end{figure}

As it can be seen in~\ref{fig:static_movie_stats}, the EVPA and fractional polarization is correctly recovered, without time evolution bias.

\end{document}